\documentclass[
preprint,
amsmath,amssymb, aps, prd, showpacs, nofootinbib ]{revtex4}
\usepackage{graphicx}
\usepackage{dcolumn}
\usepackage{bm}
\usepackage{subfigure}
\usepackage{amssymb, graphics}
\usepackage{slashed}
\usepackage[hypertex]{hyperref}
\usepackage[mathlines]{lineno}
\usepackage{epsfig}

\def\cO{\mathcal{O}}
\def\cP{\mathcal{P}}
\def\bq{\mathbf{q}}
\def\Oasv{O(\alpha_{s}v^{2})}
\def\msb{{\overline{\text{MS}}}}

\newcommand{\nn}{\nonumber}
\newcommand{\bseq}{\begin{subequations}}
\newcommand{\eseq}{\end{subequations}}
\newcommand{\beq}{\begin{equation}}
\newcommand{\eeq}{\end{equation}}
\newcommand{\bqa}{\begin{eqnarray}}
\newcommand{\eqa}{\end{eqnarray}}

\newcommand{\mcdot}{\!\cdot\!}

\newcommand{\state}[3]{^{#1}\!#2_{#3}}
\newcommand{\cstate}[4]{^{#1}\!#2_{#3}^{[#4]}}

\begin{document}
\title{QCD and relativistic $O(\alpha_{s}v^2)$ corrections to
hadronic decays of spin-singlet heavy quarkonia $h_c, h_b$ and
$\eta_b$}
\author{Jin-Zhao Li}
\email{lijinzhao86@gmail.com} \affiliation{Department of Physics and
State Key Laboratory of Nuclear Physics and Technology,
             \\Peking  University, Beijing 100871, China}
\author{Yan-Qing Ma}
\email{yqma@bnl.gov} \affiliation{Physics Department, Brookhaven
National Laboratory, Upton, New York 11973, USA}
\author{Kuang-Ta Chao}
\email{ktchao@pku.edu.cn} \affiliation{Department of Physics and
State Key Laboratory of Nuclear Physics and Technology, and Center
for High Energy Physics,
\\Peking  University, Beijing 100871, China}

\abstract{We calculate the annihilation decay widths of spin-singlet
heavy quarkonia $h_c, h_b$ and $\eta_b$} into light hadrons with
both QCD and relativistic corrections at order $O(\alpha_{s}v^{2})$
in nonrelativistic QCD. With appropriate estimates for the
long-distance matrix elements by using the potential model and
operator evolution method, we find that our predictions of these
decay widths are consistent with recent experimental measurements.
We also find that the $O(\alpha_{s}v^{2})$ corrections are small for
$b\overline{b}$ states but substantial for $c\overline{c}$ states.
In particular, the negative contribution of $O(\alpha_{s}v^{2})$
correction to the $h_{c}$ decay can lower the decay width, as
compared with previous predictions without the $O(\alpha_{s}v^{2})$
correction, and thus result in a good agreement with the recent BESIII
measurement.}

\maketitle

\section{INTRODUCTION}
The inclusive annihilation decay of heavy quarkonium is one of the
important issues in heavy quarkonium physics. It is widely accepted
that the heavy quarkonium inclusive annihilation decay can be
described by nonrelativistic QCD (NRQCD)
factorization~\cite{PhysRevD.51.1125}. In this framework, the
long-distance effects that cannot be calculated perturbatively are
described by the long-distance matrix elements (LDMEs), which are
classified in the order of $v$, the relative velocity of heavy
quarks in quarkonium. As $v$ is small in heavy quarkonium system, we
need to keep only a few number of LDMEs in the calculation. Recently,
more precise measurements for heavy quarkonium decay widths and
branching ratios are available
~\cite{Rubin:2005px,Andreotti:2005vu,Dobbs:2008ec,Ablikim:2010rc,
CLEO:2011aa,Mizuk:2012pb,:2008vj,:2009pz,Bonvicini:2009hs,Adachi:2011ji,Ge:2011kq}.
Thus, it is necessary to provide more precise theoretical
predictions to compare with the data.

For charmonium, the $c\overline{c}$ system, the inclusive
annihilation hadronic decay (into gluons and light quark pairs)
widths for $S$-, $P$-, and $D$-wave states are all calculated
up to $O(\alpha_{s})$ in
NRQCD~\cite{huang:199601,pertrelli96,huang199606,Petrelli:1997ge,
Maltoni:2000km,Fan:2009cj,He:2009bf}. Particularly, for the $S$-wave
state $\eta_c$, the $O(\alpha_{s}v^{2})$ corrections have recently
been carried out \cite{PhysRevD.83.114038}, which means the
short-distance coefficients of $O(v^{2})$ LDMEs are calculated
perturbatively to next-to-leading order (NLO) in $\alpha_{s}$. After
taking the $O(\alpha_{s}v^{2})$ corrections into account, the
measurements of $\eta_c$ decay can be described much better in
NRQCD. For the $P$-wave state $h_{c}$,  the earlier theoretical
result at $O(\alpha_{s})$ predicts the hadronic decay width of
$h_{c}$ to be about $0.72$ MeV\cite{Maltoni:2000km}, which is a
factor of 2 larger than the latest measurements by BESIII, where the
central value of the total width is about $0.73$ MeV and the
hadronic decay branching ratio is about $50\%$\cite{Ablikim:2010rc}.
Thus it is needed to study higher order in $v$ corrections to
examine whether the gap between theoretical predictions and
experimental measurements can be explained. It will be an
interesting test for the validity of NRQCD factorization for
charmonium system.

For bottomonium, the $b\overline{b}$ system, the value of $v^{2}$ is
about $0.1$, which is much smaller than $v^{2}\approx 0.3$ for
charmonium. It is then expected that the $v^{2}$ expansion should be
better for bottomonium, thus the study of bottomonium is more solid
to check NRQCD factorization.  Recently, the process $h_{b}(1P)
\rightarrow \eta_{b}(1S)\gamma$ was measured by the Belle
Collaboration~\cite{Mizuk:2012pb}. It was found that the $\eta_{b}$
decay width was about $12.4$ MeV and the decay branching fraction
of $\mathcal{B}[h_{b}(1P) \rightarrow
\eta_{b}(1S)\gamma]=49.2\pm5.7^{+5.6}_{-3.3}\%$. It is tempting to
try to explain these data in NRQCD.

In this paper, we will perform the $O(\alpha_{s}v^{2})$ calculations
for the spin-singlet $P$-wave charmonium $h_{c}$ and bottomonium
$h_{b}$, and also for the spin-singlet $S$-wave bottomonium
$\eta_{b}$. We find these corrections are important to understand
the measured data. The rest of this paper is organized as follows.
In Sec.~\ref{sec:NRQCD} we briefly introduce the NRQCD factorization
formulism in heavy quarkonium annihilation decays. Then we describe
some technical method in calculating $O(\alpha_{s}v^{2})$
short-distance coefficients in Sec.\ref{sec:method}. The results for
$S$-wave and $P$-wave states including real and virtual
contributions are presented in Sec.~\ref{sec:SDresults}. With these
results and appropriate estimates of the LDMEs, we discuss the
related phenomenology  in Sec.~\ref{sec:Phenomenology}. In the
Appendix~\ref{NRQCDIR}, we calculate the evolution of LDMEs at
$\Oasv$. In the Appnedix~\ref{scheme}, we describe our factorization
scheme choice and show how to eliminate higher twist operators.
Finally, we give a brief summary in Sec.~\ref{sec:summary}.

\section{NRQCD FACTORIZATION FOR QUARKONIUM DECAY}\label{sec:NRQCD}

In this section, we introduce the NRQCD factorization formula for
the rates of spin-singlet heavy quarkonium ($\eta_{c,b}$ and
$h_{c,b}$) decays to light hadrons. The inclusive annihilation decay
width of heavy quarkonium can be factorized by the following
formula~\cite{PhysRevD.51.1125}
\begin{eqnarray}
\Gamma(H)=\sum_{n} \frac{2 \textrm{ Im}f_{n}(\mu_{\Lambda})}{m_{Q}^{d_{n}-4}}\langle H|\cO_{n}(\mu_{\Lambda})|H\rangle,\label{NRQCD factorization}
\end{eqnarray}
where $\textrm{Im}f_{n}(\mu_{\Lambda})$ is the short-distance (SD)
coefficient that can be perturbatively calculated using full QCD
Lagrangian. The LDMEs $\langle
H|\cO_{n}(\mu_{\Lambda})|H\rangle$ involve non-perturbative effects
and are classified by the relative velocity $v$ between $Q$ and
$\overline{Q}$, according to power counting in
Refs.~\cite{PhysRevD.51.1125,Brambilla:1999xf,PhysRevD.63.054007,PhysRevD.64.036002,Brambilla:2008zg}.

The NRQCD Lagrangian can be derived by integrating out the degrees
of freedom of order $m_{Q}$, the mass of the heavy quark, from the QCD
Lagrangian, which gives
\begin{eqnarray}
\mathcal{L}_{\textrm{NRQCD}} = \mathcal{L}_{\textrm{light}}+\mathcal{L}_{\textrm{heavy}}+\delta\mathcal{L}.
\end{eqnarray}
The heavy part of the Lagrangian describes the motions of
(anti-)heavy quark in spacetime and is given by
\begin{eqnarray}\label{kineticLag}
\mathcal{L}_{\textrm{heavy}} = \psi^{\dagger} (i D_{t}+\frac{\mathbf{D}^2}{2m_{Q}}) \psi+\chi^{\dagger} (i D_{t}-\frac{\mathbf{D}^2}{2m_{Q}}) \chi
\end{eqnarray}
where $\psi$($\chi$) denotes the Pauli spinor field that annihilates
(creates) a heavy (anti-)quark, and $D_{t}$($\mathbf{D}$) is the
time(space) component of the gauge-covariant derivative $D^{\mu}$.
The light piece of the Lagrangian reads
\begin{eqnarray}
\mathcal{L}_{\textrm{light}} = -\frac{1}{2}\textrm{Tr }G^{\mu\nu}G_{\mu\nu}+\sum_{n_{f}}\overline{q}i\slashed{D}q
\end{eqnarray}
where $G^{\mu\nu}$ is the gluon field strength tensor, $q$ is the
Dirac spinor field of light quarks and $n_{f}$ is the number of
light flavors. The bilinear Lagrangian term which contains the order
$v^{2}$ correction is
\begin{eqnarray}\label{bilinearLag}
\delta\mathbf{L}_{\textrm{bilinear}}&=&\frac{c_{1}}{8 m_{Q}^{3}}\psi^{\dagger}
(\mathbf{D}^{2})^{2}\psi + \frac{c_{2}}{8 m_{Q}^{2}}\psi^{\dagger}(\mathbf{D}\cdot g\mathbf{E}-g\mathbf{E}\mcdot\mathbf{D})\psi\nonumber\\
&+&\frac{c_{3}}{8 m_{Q}^{2}}\psi^{\dagger}(i\mathbf{D}\!\times\! g\mathbf{E}-
g\mathbf{E}\!\times\! i\mathbf{D})\mcdot\mathbf{\sigma}\psi + \frac{c_{4}}{2 m_{Q}}\psi^{\dagger}(g\mathbf{B}\mcdot \mathbf{\sigma})\psi\nonumber\\
&+&\textrm{charge conjugate terms},
\end{eqnarray}
where $E^{i}=G^{0i}$ and $B^{i}=\frac{1}{2}\epsilon^{ijk}G^{jk}$ are
the electric and magnetic components of the gluon field strength
tensor $G^{\mu\nu}$, and $c_{i}=1+O(\alpha_{s}),i=1,2,3,4$ are the
dimensionless coefficients corresponding to each operator.

In order to describe the annihilation decay of quarkonium, a set of
local four-fermion operators $\cO_{i}$ which appear in Eq.~(\ref{NRQCD
factorization}) are needed. For example, the operator
$\psi^{\dagger}\chi\chi^{\dagger}\psi$ can annihilate a
$Q\overline{Q}$ pair in the $\cstate{1}{S}{0}{1}$ configuration.  In
our case, for the $O(\alpha_{s}v^{2})$ calculation of spin-singlet
quarkonium decay, the power counting rules~\cite{PhysRevD.51.1125}
give the following seven operators and LDMEs in Eq.~(\ref{NRQCD
factorization}): for $S$-wave quarkonium,
 \bseq
\begin{eqnarray}
\cO(^{1}S_{0}^{[1]})&=&\psi^{\dagger}\chi\chi^{\dagger}\psi,\\
\mathcal{P}(^{1}S_{0}^{[1]})&=&\frac{1}{2}\psi^{\dagger}\chi\chi^{\dagger}(-\frac{i\tensor{\mathbf{D}}}{2})^{2}\psi+\textrm{h.c.},
\end{eqnarray}
\eseq
 for $P$ wave quarkonium,
 \bseq
\begin{eqnarray}
\cO(^{1}S_{0}^{[8]})&=&\psi^{\dagger}T^{a}\chi\chi^{\dagger}T^{a}\psi,\\
\mathcal{P}(^{1}S_{0}^{[8]})&=&\frac{1}{2}\psi^{\dagger}T^{a}\chi\chi^{\dagger}T^{a}(-\frac{i\tensor{\mathbf{D}}}{2})^{2}\psi+\textrm{h.c.},\\
\cO(^{1}P_{1}^{[1]})&=&\psi^{\dagger}(-\frac{i\tensor{\mathbf{D}}}{2})\chi\mcdot\chi^{\dagger}(-\frac{i\tensor{\mathbf{D}}}{2})\psi,\\
\mathcal{P}(^{1}P_{1}^{[1]})&=&\frac{1}{2}\psi^{\dagger}
(-\frac{i\tensor{\mathbf{D}}}{2})\chi\mcdot\chi^{\dagger}(-\frac{i\tensor{\mathbf{D}}}{2})^{3}\psi+\textrm{h.c.},\\
\mathcal{T}_{1-8}(\state{1}{S}{0},\state{1}{P}{1})&=&\frac{1}{2}\psi^\dagger
g\mathbf{E}\chi\cdot\chi^\dagger\tensor{\mathbf{D}}\psi+\textrm{h.c.}\label{eq:t18},
\end{eqnarray}
\eseq
and
\bseq
\begin{eqnarray}
\langle \cO(\cstate{2S+1}{L}{J}{1,8})\rangle_{H} &\equiv& \langle H|\cO(\cstate{2S+1}{L}{J}{1,8})| H\rangle,\\
\langle \mathcal{P}(\cstate{2S+1}{L}{J}{1,8})\rangle_{H} &\equiv& \langle H|\mathcal{P}(\cstate{2S+1}{L}{J}{1,8})| H\rangle.
\end{eqnarray}
\eseq
 Note that, choosing different power counting rules, one may
get a different set of operators. For example, in the power counting
rule of Ref.~\cite{Brambilla:2008zg}, $m_{Q}$ and $v$ are
homogeneous, which gives that the chromomagnetic field $g\mathbf{B}$
scales as $(m_{Q}v)^2$. While that field scales as $m_{Q}^{2}v^4$ in
Ref.~\cite{PhysRevD.51.1125}, which is further suppressed by $v^2$.
As a result, many operators considered in
Ref.~\cite{Brambilla:2008zg} disappear in our calculation, leaving
the above seven. These seven matrix elements are all independent
with each other, i.e. they cannot be eliminated by field
redefinition or Poincare invariance~\cite{Brambilla:2008zg}.

Using the seven operators, we give the explicit form of
Eq.~(\ref{NRQCD factorization}) for $\state{1}{S}{0}$ and
$\state{1}{P}{1}$ states,
 \bseq \bqa
\Gamma(H(\state{1}{S}{0})\rightarrow
\textrm{LH})=\frac{F(\cstate{1}{S}{0}{1})}{m_{Q}^{2}}\langle
\cO(\cstate{1}{S}{0}{1})\rangle_{\state{1}{S}{0}}+\frac{G(\cstate{1}{S}{0}{1})}{m_{Q}^{4}}\langle
\mathcal{P}(\cstate{1}{S}{0}{1})\rangle_{\state{1}{S}{0}},\label{1S0
factorization}
 \eqa
\begin{eqnarray}
\Gamma(H(\state{1}{P}{1})\rightarrow
\textrm{LH})&=&\frac{F(\cstate{1}{S}{0}{8})}{m_{Q}^{2}}\langle
\cO(\cstate{1}{S}{0}{8})\rangle_{\state{1}{P}{1}}+\frac{G(\cstate{1}{S}{0}{8})}{m_{Q}^{4}}\langle
\mathcal{P}(\cstate{1}{S}{0}{8})\rangle_{\state{1}{P}{1}}\nonumber\\
&+&\frac{F(\cstate{1}{P}{1}{1})}{m_{Q}^{4}}\langle
\cO(\cstate{1}{P}{1}{1})\rangle_{\state{1}{P}{1}}+
\frac{G(\cstate{1}{P}{1}{1})}{m_{Q}^{6}}\langle
\mathcal{P}(\cstate{1}{P}{1}{1})\rangle_{\state{1}{P}{1}}.\label{1P1
factorization}
\end{eqnarray}
\eseq
 Note that, we omit a term of $\frac{T(\state{1}{S}{0},\state{1}{P}{1})}{m_{Q}^{5}}\langle \mathcal{T}_{1-8}(\state{1}{S}{0},
\state{1}{P}{1})\rangle_{\state{1}{P}{1}}$ in Eq.~\eqref{1P1
factorization} to simplify our theoretical framework, although the
LDME $\langle
\mathcal{T}_{1-8}(\state{1}{S}{0},\state{1}{P}{1})\rangle_{\state{1}{P}{1}}$
is of the same order  in $v$ as $\langle
\mathcal{P}(\cstate{1}{P}{1}{1})\rangle_{\state{1}{P}{1}}$. There
are two reasons that lead us to do this simplification. Numerically,
this contribution is small, which is because
$T(\state{1}{S}{0},\state{1}{P}{1})$ vanishes at leading order (LO)
in $\alpha_s$ due to the charge parity conservation. Theoretically,
and more importantly, this contribution is finite, that is, no
infrared (IR) poles are needed to cancel between this channel and
other four channels in Eq.~\eqref{1P1 factorization}. It is then
impossible to distinguish this finite contribution from the
renormalization scheme or factorization scheme choice of other
operators, such as
$\langle\mathcal{O}(\cstate{1}{P}{1}{1})\rangle_{\state{1}{P}{1}}$
or
$\langle\mathcal{O}(\cstate{1}{S}{0}{8})\rangle_{\state{1}{P}{1}}$.
Therefore, by ignoring this operator in the hadronic decay width, it
is equivalent that we choose a specific renormalization scheme or
factorization scheme for other operators. In Appendix~\ref{scheme},
we will give an explicit definition of our factorization scheme to
absorb the term
$\frac{T(\state{1}{S}{0},\state{1}{P}{1})}{m_{Q}^{5}}\langle
\mathcal{T}_{1-8}(\state{1}{S}{0},
\state{1}{P}{1})\rangle_{\state{1}{P}{1}}$. Although our scheme is
in principle distinguished from $\msb$ scheme, as we will discussed
in Appendix~\ref{scheme}, there is no difference between these two
schemes for our purpose in this work. As a result, we will pretend
to use $\msb$ scheme in the following.

Through the above factorization formula, one can match full QCD with
NRQCD to get the short-distance (SD) coefficients $F$ and $G$
perturbatively. The skeleton of the matching procedure is given by
\begin{eqnarray}\label{matching}
\textrm{Im}\mathcal{M}(Q\overline{Q}\rightarrow
Q\overline{Q})\Big{|}_{\textrm{pert QCD}}=\sum_{n} \frac{2 \textrm{
Im}f_{n}(\mu_{\Lambda})}{m_{Q}^{d_{n}-4}}\langle Q\overline{Q} |
\cO_{n}(\mu_{\Lambda}) |Q\overline{Q}\rangle
\Big{|}_{\textrm{NRQCD}},
\end{eqnarray}
The determination of SD coefficients will be discussed in detail in
the next section.

\section{DETAILS IN FULL QCD CALCULATION}\label{sec:method}

\subsection{Kinematics}
We work in the rest frame of the heavy quarkonium. It is customary
to decompose the momenta of $Q$ and $\overline{Q}$ as
\begin{subequations}
\begin{eqnarray}
p_{Q}& = &\frac{1}{2}P+q,\\
p_{\overline{Q}}& = &\frac{1}{2}P-q,
\end{eqnarray}
\end{subequations}
where $P$ is the total momentum and $q$ is half of the relative
momentum, which satisfies the relation $P\mcdot q = 0$. The explicit
four-vector form of $P$ and $q$ in the rest frame are
 \bseq \bqa
P& = &(2E_{\bq},\mathbf{0}),\\
q& = &(0,\bq), \eqa \eseq
with $E_{\bq} = \sqrt{m_{Q}^{2}+\bq^{2}}$.

The treatment of final state phase space integration at $\Oasv$
level is slightly different from ordinary calculations (i.e. leading
order of $v$ calculation). To make it simpler, we use the following
rescaling transformation for all external
momenta~\cite{PhysRevD.66.094011,PhysRevD.83.114038}, \bseq \bqa
P &\rightarrow& P^{\prime}\frac{E_{\bq}}{m_{Q}},\\
k_{f} &\rightarrow& k_{f}^{\prime}\frac{E_{\bq}}{m_{Q}}, \eqa \eseq
but keep the relative momentum $q$ and loop integral momentum $l$
unchanged. Once we take such trick, the $\bq^{2}$ dependence in both
phase space and current factor [i.e. $1/(2M)$ where $M$ is the
quarkonium mass] can be absorbed into the amplitude, then we can safely
take $\bq \rightarrow 0$ in these terms and only expand $\bq,\bq^\prime$ at the
amplitude level, where $\bq^\prime$ is half of the relative momentum between
$Q\overline{Q}$ pair on the complex conjugate side. (Note that $|\bq|=|\bq^\prime|$
but their direction does not need to be the same, so in general $\bq\ne\bq^\prime$).
It should be kept in mind that this
trick can only work in the case where all final state partons are
massless (i.e. gluons and light quarks), because, in the case of massive
partons, the on-shell relation does not hold under rescaling,
which will break the QCD gauge invariance.

\subsection{Covariant Projection Method in D-Dimension}

Instead of using matching method directly, we use an equivalent but
more efficient method, i.e., the covariant projection method, to
calculate the imaginary part of the SD coefficients in Eqs.~(\ref{1S0
factorization}) and (\ref{1P1 factorization}). In order to get
spin-singlet $Q\overline{Q}$ decay amplitudes, we take the following
spin and color projectors onto $Q\overline{Q}$ quark
lines~\cite{PhysRevD.27.1518}:
 \bqa \Pi_{0} =
\frac{1}{2\sqrt{2}(E_{\bq}+m_{Q})}(\frac{\slashed{P}}{2}+\slashed{q}+m_{Q})\frac{(\slashed{P}+2
E_{\bq})\gamma_{5}(-\slashed{P}+2 E_{\bq})}{8
E_{\bq}^2}(\frac{\slashed{P}}{2}-\slashed{q}-m_{Q}),\label{spin0projector}
\eqa
 and
 \bseq \bqa
\mathcal{C}_{1} &=& \frac{\mathbf{1}}{\sqrt{N_{c}}},\\
\mathcal{C}_{8} &=& \sqrt{2}\mathbf{T}^{a}. \eqa \eseq
 We do Taylor expansion of the projected amplitudes in powers of $q$ to the required order,
 \bqa
\mathcal{M}(q)& = &\mathcal{M}(0)+\frac{\partial
\mathcal{M}(q)}{\partial q^{\alpha}}\Big{|}_{q=0}
q^{\alpha}+\frac{1}{2!}\frac{\partial^{2} \mathcal{M}(q)}{\partial
q^{\alpha}
\partial q^{\beta}}\Big{|}_{q=0} q^{\alpha}q^{\beta}\nn\\
& + &\frac{1}{3!}\frac{\partial^{3} \mathcal{M}(q)}{\partial q^{\alpha}
\partial q^{\beta} \partial q^{\gamma}}\Big{|}_{q=0} q^{\alpha}q^{\beta}q^{\gamma}+\cdots,
\eqa
 and then make the replacement:
 \bseq \bqa
q_{\alpha}q_{\beta}& \rightarrow &\frac{\bq^{2}}{D-1}\Pi_{\alpha\beta},\label{q2rep}\\
q_{\alpha}q_{\beta}^{\prime}& \rightarrow &\frac{\bq\mcdot\bq^{\prime}}{D-1}\Pi_{\alpha\beta},\label{q12rep}\\
q_{\alpha}q_{\beta}q_{\gamma}q_{\lambda}^{\prime}& \rightarrow
&\frac{\bq^{2}\bq\mcdot\bq^{\prime}}{D+1}(\Pi_{\alpha\beta}\Pi_{\gamma\lambda}+
\Pi_{\alpha\gamma}\Pi_{\beta\lambda}+\Pi_{\alpha\lambda}\Pi_{\gamma\beta}),\label{q4rep}
\eqa \eseq
 to project them to definite states, where
 \beq \Pi_{\alpha\beta} =
-g_{\alpha\beta}+\frac{P_{\alpha}^{\prime}P_{\beta}^{\prime}}{4
m_{Q}^{2}}, \eeq
with $P^{\prime}$ the rescaled heavy quarkonium
momentum. For example, the third derivative term of $\mathcal{M}$
convolutes with the first derivative term of $\mathcal{M}^{\dagger}$
giving the squared amplitudes term,
 \bqa
&&\frac{1}{3!}\frac{\partial^{3} \mathcal{M}(q)}{\partial q^{\alpha}
\partial q^{\beta} \partial q^{\gamma}}\Big{|}_{q=0} \frac{\partial
\mathcal{M}^{\dagger}(q^{\prime})}{\partial q^{\prime\lambda} }\Big{|}_{q^{\prime}=0} q^{\alpha}q^{\beta}q^{\gamma}q^{\prime\lambda}\nn\\
& \rightarrow &
\frac{1}{3!}\frac{\bq^{2}\bq\mcdot\bq^{\prime}}{D+1}(\Pi_{\alpha\beta}
\Pi_{\gamma\lambda}+\Pi_{\alpha\gamma}\Pi_{\beta\lambda}+\Pi_{\alpha\lambda}\Pi_{\gamma\beta})\frac{\partial^{3}
\mathcal{M}(q)}{\partial q^{\alpha} \partial q^{\beta} \partial
q^{\gamma}}\Big{|}_{q=0} \frac{\partial
\mathcal{M}^{\dagger}(q^{\prime})}{\partial q^{\prime\lambda}
}\Big{|}_{q^{\prime}=0} \eqa
 which contributes to the SD
coefficient of $G(^{1}P_{1}^{[1]})$ in Eq. (\ref{1P1
factorization}).

\section{PERTURBATIVE QCD RESULTS OF SHORT-DISTANCE COEFFICIENTS}\label{sec:SDresults}
We generate Feynman diagrams and amplitudes by {\sf
FeynArts}~\cite{Mertig1991,Hahn:2000kx}, and then calculate the
squared amplitudes by self-written {\sf Mathematica} codes. The
phase space integrals are done analytically using the method
presented in Ref.~\cite{Petrelli:1997ge}. Ultra-violet(UV) and
IR divergences are both regularized by dimensional
regularization. The renormalizations for heavy quark mass $m_{Q}$,
heavy quark field $\psi_{Q}$, light quark field $\psi_{q}$ and gluon
field $A_{\mu}$ are in the on-mass-shell scheme(OS), and that for
the QCD coupling constant $g_{s}$ is in the $\overline{MS}$ scheme,
\bseq \bqa
\delta Z_{m_{Q}}^{OS} &=& -3 C_{F}\frac{\alpha_{s}}{4\pi}N_{\epsilon}\left[\frac{1}{\epsilon_{UV}}+\frac{4}{3}\right],\\
\delta Z_{2}^{OS} &=& -C_{F}\frac{\alpha_{s}}{4\pi}N_{\epsilon}\left[\frac{1}{\epsilon_{UV}}+\frac{2}{\epsilon_{IR}}+4\right],\\
\delta Z_{2l}^{OS} &=& -C_{F}\frac{\alpha_{s}}{4\pi}N_{\epsilon}\left[\frac{1}{\epsilon_{UV}}-\frac{1}{\epsilon_{IR}}\right],\\
\delta Z_{3}^{OS} &=& \frac{\alpha_{s}}{4\pi}N_{\epsilon}\left[(\beta_{0}-2C_{A})\left(\frac{1}{\epsilon_{UV}}-\frac{1}{\epsilon_{IR}}\right)\right],\\
\delta Z_{g}^{\overline{MS}} &=&
-\frac{\beta_{0}}{2}\frac{\alpha_{s}}{4\pi}N_{\epsilon}\left[\frac{1}{\epsilon_{UV}}+\ln\frac{m_{Q}^{2}}{\mu_{r}^{2}}\right],
\eqa \eseq where
$N_{\epsilon}(m_{Q})=(\frac{4\pi\mu_{r}^{2}}{m_{Q}^{2}})^{\epsilon}\Gamma(1+\epsilon)$
is an overall factor, and $\mu_{r}$ is the renormalization scale.
$\beta_{0}=\frac{11}{3}C_{A}-\frac{4}{3}T_{F}n_{f}$ is the one-loop
coefficient of the $\beta$ function, $n_f$ is the active quark
flavors, which we set to be 3 for charmonium and 4 for bottomonium.

\subsection{Short-Distance Coefficients of S-Wave Quarkonium Hadronic Decay}
\begin{figure}
\includegraphics[scale=1.3]{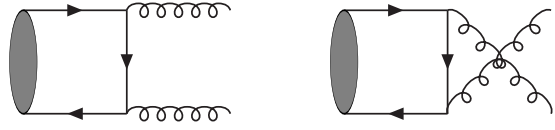}
\caption{\label{borndiagram} Born level Feynman diagrams for
$\cstate{1}{S}{0}{1},\cstate{1}{S}{0}{8} \rightarrow gg$.}
\end{figure}

Leading order in $\alpha_{s}$ calculations give the Born level decay
width and its relativistic correction, respectively, as
\bseq\label{Gamma1S01born} \bqa \Gamma_{\mbox{{\footnotesize
Born}}}(\cstate{1}{S}{0}{1} \rightarrow gg) &=& \frac{4}{3}
(4\pi\alpha_{s})^{2}\frac{\mu_{r}^{4\epsilon}}{m_{Q}^{2}}
\Phi_{(2)}(1-\epsilon)(1-2\epsilon)\frac{\langle\cO(\cstate{1}{S}{0}{1})\rangle_{\state{1}{S}{0}}^{\mbox{{\footnotesize Born}}}}{2 N_{c}},\\
\Gamma_{\mbox{{\footnotesize Born}}}^{(v^{2})}(\cstate{1}{S}{0}{1}
\rightarrow gg) &=&
-\frac{2(2-\epsilon)}{3-2\epsilon}\frac{\bq^{2}}{m_{Q}^{2}}\Gamma_{\mbox{{\footnotesize
Born}}}(\cstate{1}{S}{0}{1} \rightarrow gg), \eqa \eseq where
$\Phi_{(2)}=\frac{1}{8\pi}(\frac{4\pi}{M^{2}})^{\epsilon}\frac{\Gamma(1-\epsilon)}{\Gamma(2-2\epsilon)}$
is the total two-body phase space in $D$ dimension and $M =
2m_{Q}\sqrt{1+\frac{\bq^{2}}{m_{Q}^{2}}}$ is the quarkonium mass
including the relativistic correction. The two Born diagrams are
illustrated in Fig.~\ref{borndiagram}.

\begin{figure}
\includegraphics[scale=1.3]{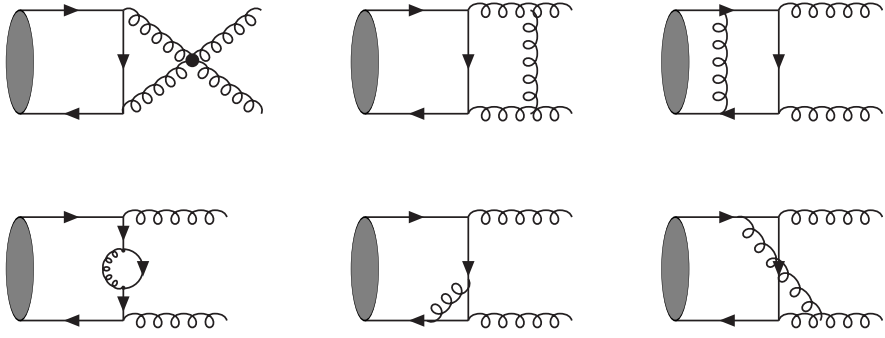}
\caption{\label{loopdiagram} Virtual correction Feynman diagrams for
$\cstate{1}{S}{0}{1},\cstate{1}{S}{0}{8} \rightarrow gg$. The crossed diagrams have been suppressed.}
\end{figure}

The next-to-leading order calculations include real and virtual
corrections. For $S$-wave Fock states (i.e. $\cstate{1}{S}{0}{1}$ and
$\cstate{1}{S}{0}{8}$), UV divergences will be canceled by
counterterm diagrams, and most IR divergences will be canceled
between real and virtual corrections, leaving some residue
divergences at $O(v^2)$. The cancelation of such residue divergences
will be presented in the next section by calculating NRQCD LDMEs at
one-loop level. The contribution of virtual plus counterterm
corrections is \bqa\label{Gamma1S01virtual}
\begin{split}
\Gamma_{\mbox{{\footnotesize Virtual}}}(\cstate{1}{S}{0}{1} \rightarrow gg) =&
\frac{3\alpha_{s}}{\pi}\Gamma_{\mbox{{\footnotesize Born}}}(\cstate{1}{S}{0}{1}
\rightarrow gg)f_{\epsilon}(m_{Q})\Big\{[-\frac{1}{\epsilon^{2}}-\frac{1}{6}\beta_{0}\frac{1}{\epsilon}\\
&+\frac{1}{36}(-6\beta_{0}\ln(\frac{4 m_{Q}^{2}}{\mu_{r}^{2}})+19\pi^{2}-44)]\\
&+\frac{\bq^{2}}{m_{Q}^{2}}[\frac{4}{3}\frac{1}{\epsilon^{2}}-\frac{4n_{f}-97}{27}\frac{1}{\epsilon}\\
&-\frac{1}{324}(-72\beta_{0}\ln(\frac{4m_{Q}^{2}}{\mu_{r}^{2}})+8n_{f}+267\pi^{2}-280)]\Big\},
\end{split}
\eqa
where $f_{\epsilon}(m_{Q})=(\frac{\pi\mu_{r}^{2}}{m_{Q}^{2}})^{\epsilon}\Gamma(1+\epsilon)$.
Some selected Feynman diagrams are shown in Fig.~\ref{loopdiagram}.

\begin{figure}
\includegraphics[scale=1.3]{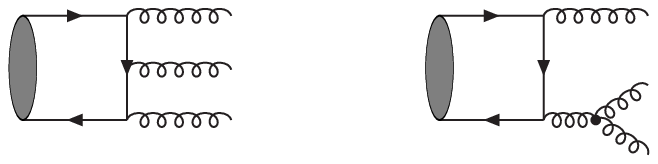}
\caption{\label{realgggdiagram} Real correction Feynman diagrams for
$\cstate{1}{S}{0}{1}, \cstate{1}{S}{0}{8},\cstate{1}{P}{1}{1} \rightarrow ggg$. The crossed diagrams have
been suppressed. The second diagram vanishes in $\cstate{1}{P}{1}{1}$.}
\includegraphics[scale=1.3]{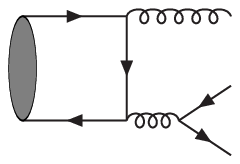}
\caption{\label{realqqgdiagram} Real correction Feynman diagrams for
$\cstate{1}{S}{0}{1}, \cstate{1}{S}{0}{8} \rightarrow q\overline{q}g$. The crossed diagrams have
been suppressed.}
\end{figure}

The real correction contains two sets, where one set is the final
states with $ggg$ and the other one with $q\overline{q}g$. Some
typical Feynman diagrams are shown in Fig.~\ref{realgggdiagram} and
Fig.~\ref{realqqgdiagram} and the contributions to decay width are
\bseq\label{Gamma1S01real}
\begin{align}
\begin{split}
\Gamma(\cstate{1}{S}{0}{1} \rightarrow ggg) =&\frac{3\alpha_{s}}{\pi}
\Gamma_{\mbox{{\footnotesize Born}}}(\cstate{1}{S}{0}{1} \rightarrow gg)
f_{\epsilon}(m_{Q})\Big\{[\frac{1}{\epsilon^{2}}+\frac{11}{6}\frac{1}{\epsilon}+\frac{1}{72}(724-69\pi^{2})]\\
&+\frac{\bq^{2}}{m_{Q}^{2}}[-\frac{4}{3}\frac{1}{\epsilon^2}-\frac{3}{\epsilon}-\frac{437-42\pi^{2}}{27}]\Big\},
\end{split}\\
\Gamma(\cstate{1}{S}{0}{1} \rightarrow q\overline{q}g) =& \frac{n_{f}}{2}\frac{\alpha_{s}}{\pi}
\Gamma_{\mbox{{\footnotesize Born}}}(\cstate{1}{S}{0}{1} \rightarrow gg)
\frac{f_{\epsilon}(m_{Q})}{\Gamma(1+\epsilon)\Gamma(1-\epsilon)}[-\frac{2}{3}
\frac{1}{\epsilon}-\frac{16}{9}+\frac{\bq^{2}}{m_{Q}^{2}}(\frac{8}{9}\frac{1}{\epsilon}+\frac{86}{27})].
\end{align}
\eseq Combining Eqs.~(\ref{Gamma1S01born}),~(\ref{Gamma1S01virtual})
and~(\ref{Gamma1S01real}), we obtain the hadronic decay width with
both QCD radiative and relativistic corrections at NLO of
$\state{1}{S}{0}$ heavy quarkonium, \bqa\label{Gamma1S0noNRQCD}
\begin{split}
\Gamma_{\mbox{\footnotesize QCD}}(\state{1}{S}{0} \rightarrow \mbox{LH})=&
\Gamma_{\mbox{{\footnotesize Born}}}(\cstate{1}{S}{0}{1} \rightarrow gg)
\Big\{\big[1+\frac{\alpha_s}{\pi}f_{\epsilon}(m_{Q})\frac{1}{72}(-36\beta_{0}\ln(\frac{4m_{Q}^{2}}{\mu_{r}^{2}})\\
&-64n_{f}-93\pi^2+1908)\big]-\frac{4}{3}\frac{\bq^2}{m_{Q}^{2}}\big[1+\frac{\alpha_s}{\pi}f_{\epsilon}(m_{Q})(-\frac{4}{3}\frac{1}{\epsilon}\\
&+\frac{1}{144}(-72\beta_{0}\ln(\frac{4m_{Q}^{2}}{\mu_{r}^{2}})-164n_{f}-237\pi^{2}+4964))\big]\Big\}.
\end{split}
\eqa We note that our results agree with the previous work for
$O(\alpha_{s}v^2)$ correction~\cite{PhysRevD.83.114038} and
$O(\alpha_{s})$ correction~\cite{Petrelli:1997ge,Fan:2009cj}.
Comparing our results with Ref.~\cite{PhysRevD.83.114038}, a slight
difference of two body phase space $\Phi_{2}$ between them can be
found. In Ref.~\cite{PhysRevD.83.114038} $\Phi_{2}$ is defined so as
to remove the $\bq^{2}$ dependence into the coefficients, so our
individual virtual and real parts, Eqs.~(\ref{Gamma1S01virtual}) and
(\ref{Gamma1S01real}), look different from the results in
Ref.~\cite{PhysRevD.83.114038}, but essentially they are equivalent.
The total NLO result Eq.~(\ref{Gamma1S0noNRQCD}) is explicitly the
same, independent of the definition of $\Phi_{2}$. The correct
repetition of the hadronic decay SD coefficients of
$\state{1}{S}{0}$ heavy quarkonium enables us to extend discussion
from charm quark system to bottom quark system (i.e. $\eta_{b}$) and
also partly checks our codes when dealing with $P$-wave heavy
quarkonium.

\subsection{Short-Distance Coefficients of P-Wave Quarkonium Hadronic Decay}
The procedure in calculating the $\state{1}{P}{1}$ heavy quarkonium
is similar to $\state{1}{S}{0}$, although more complicated.
Additional simplification can be taken by imposing $C$ (charge)
parity conservation of QCD to constrain Feynman diagrams. A
straightforward result is that $C$ parity conservation prohibits
$\cstate{1}{P}{1}{1}$ Fock state, which has $C=-1$, to decay to two
gluons, whose $C=+1$, no matter they are real or virtual. By tedious
but straightforward calculation, we get the results as follows.

At the Born level, \bseq\label{Gamma1S08born} \bqa
\Gamma_{\mbox{{\footnotesize Born}}}(\cstate{1}{S}{0}{8} \rightarrow gg) &=& \frac{5}{12}
(4\pi\alpha_{s})^{2}\frac{\mu_{r}^{4\epsilon}}{m_{Q}^{2}}\Phi_{(2)}(1-\epsilon)(1-2\epsilon)
\langle\cO(\cstate{1}{S}{0}{8})\rangle_{\state{1}{P}{1}}^{\mbox{{\footnotesize Born}}},\\
\Gamma_{\mbox{{\footnotesize Born}}}^{(v^{2})}(\cstate{1}{S}{0}{8}
\rightarrow gg) &=&
-\frac{2(2-\epsilon)}{3-2\epsilon}\frac{\bq^{2}}{m_{Q}^{2}}\Gamma_{\mbox{{\footnotesize
Born}}}(\cstate{1}{S}{0}{8} \rightarrow gg), \eqa \eseq

For NLO corrections, \bqa \label{Gamma1S08virtual}
\begin{split}
\Gamma_{\mbox{{\footnotesize Virtual}}}(\cstate{1}{S}{0}{8} \rightarrow gg) =& \frac{3\alpha_{s}}{\pi}
\Gamma_{\mbox{{\footnotesize Born}}}(\cstate{1}{S}{0}{8} \rightarrow gg)f_{\epsilon}(m_{Q})
\Big\{[-\frac{1}{\epsilon^{2}}+\frac{n_{f}-21}{9}\frac{1}{\epsilon}\\
&+\frac{1}{72}(-12\beta_{0}\ln(\frac{4 m_{Q}^{2}}{\mu_{r}^{2}})+29\pi^{2}-16)]\\
&+\frac{\bq^{2}}{m_{Q}^{2}}[\frac{4}{3}\frac{1}{\epsilon^{2}}-\frac{4n_{f}-115}{27}\frac{1}{\epsilon}\\
&-\frac{1}{628}(-144\beta_{0}\ln(\frac{4m_{Q}^{2}}{\mu_{r}^{2}})+16n_{f}+345\pi^{2}-992)]\Big\},
\end{split}
\eqa
\bqa
\label{Gamma1S08realggg}
\begin{split}
\Gamma(\cstate{1}{S}{0}{8} \rightarrow ggg) =&\frac{3\alpha_{s}}{\pi}\Gamma_{\mbox{{\footnotesize Born}}}
(\cstate{1}{S}{0}{8} \rightarrow gg)f_{\epsilon}(m_{Q})\Big\{[\frac{1}{\epsilon^{2}}+\frac{7}{3}\frac{1}{\epsilon}-\pi^{2}+\frac{104}{9}]\\
&+\frac{\bq^{2}}{m_{Q}^{2}}[-\frac{4}{3}\frac{1}{\epsilon^2}-\frac{4}{\epsilon}-\frac{554-45\pi^{2}}{27}]\Big\},
\end{split}
\eqa
\begin{align}
\label{Gamma1S08realqqg}
\Gamma(\cstate{1}{S}{0}{8} \rightarrow q\overline{q}g) =& \frac{n_{f}}{2}
\frac{\alpha_{s}}{\pi}\Gamma_{\mbox{{\footnotesize Born}}}(\cstate{1}{S}{0}{8} \rightarrow gg)
\frac{f_{\epsilon}(m_{Q})}{\Gamma(1+\epsilon)\Gamma(1-\epsilon)}[-\frac{2}{3}\frac{1}{\epsilon}-
\frac{16}{9}+\frac{\bq^{2}}{m_{Q}^{2}}(\frac{8}{9}\frac{1}{\epsilon}+\frac{86}{27})],\\
\label{Gamma1P11realggg}
\begin{split}
\Gamma(\cstate{1}{P}{1}{1} \rightarrow ggg) =& \frac{40\alpha_{s}^{3}}{27}
f_{\epsilon}(m_{Q})(8\pi\Phi_{2})\Big\{[-\frac{1}{\epsilon}+\frac{7\pi^2}{24}-\frac{5}{3}]\\
&+\frac{\bq^{2}}{m_{Q}^{2}}[\frac{29}{15}\frac{1}{\epsilon}+\frac{4216-555\pi^{2}}{900}]
\Big\}\frac{\langle\cO(\cstate{1}{P}{1}{1})\rangle_{\state{1}{P}{1}}^{\mbox{{\footnotesize Born}}}}{2N_{c}m_{Q}^{4}},
\end{split}
\end{align}
Summing over the above results, we get the total hadronic decay
width, \bqa\label{Gamma1P1noNRQCD}
\begin{split}
\Gamma_{\mbox{\footnotesize QCD}}(\state{1}{P}{1} \rightarrow \mbox{LH})=&
\Gamma_{\mbox{{\footnotesize Born}}}(\cstate{1}{S}{0}{8} \rightarrow gg)
\Big\{\big[1+\frac{\alpha_s}{\pi}f_{\epsilon}(m_{Q})(-\frac{1}{2}\beta_{0}
\ln(\frac{4m_{Q}^{2}}{\mu_{r}^{2}})\\
&-\frac{8}{9}n_{f}-\frac{43\pi^{2}}{24}+34\big]-\frac{4}{3}\frac{\bq^2}{m_{Q}^{2}}
\big[1+\frac{\alpha_s}{\pi}f_{\epsilon}(m_{Q})(-\frac{7}{12}\frac{1}{\epsilon}\\
&+\frac{1}{288}(-144\beta_{0}\ln(\frac{4m_{Q}^{2}}{\mu_{r}^{2}})-328n_{f}-735\pi^{2}+12304))\big]\Big\}\\
&+\frac{40\alpha_{s}^{3}}{27}f_{\epsilon}(m_{Q})(8\pi\Phi_{2})\Big\{[-\frac{1}{\epsilon}+\frac{7\pi^2}{24}-\frac{5}{3}]\\
&+\frac{\bq^{2}}{m_{Q}^{2}}[\frac{29}{15}\frac{1}{\epsilon}+\frac{4216-555\pi^{2}}{900}]\Big\}
\frac{\langle\cO(\cstate{1}{P}{1}{1})\rangle_{\state{1}{P}{1}}^{\mbox{{\footnotesize Born}}}}{2N_{c}m_{Q}^{4}}.
\end{split}
\eqa

\subsection{Evaluating NRQCD LDMEs And Matching Full QCD Results}

In Eqs.~(\ref{Gamma1S0noNRQCD}) and (\ref{Gamma1P1noNRQCD}), there
exist explicit IR divergences. To cancel these divergence, we need
to evaluate LDMEs at the loop level. By replacing all the Born LDMEs
appearing in Eqs.~(\ref{Gamma1S0noNRQCD}) and (\ref{Gamma1P1noNRQCD})
by one-loop LDMEs, all IR divergences should be canceled and the
final results will be infra-red safe quantities.

The self-energy contributions that connect Born LDMEs to their
corresponding relativistic ones are first calculated in
Ref.~\cite{PhysRevD.51.1125}. The intersecting diagrams that
describe the E1 transition between $\cstate{1}{S}{0}{8}$ and
$\cstate{1}{P}{1}{1}$ states at $O(\alpha_{s}v^{2})$ in this work
are new. The detailed calculation is presented in
Appendix~\ref{NRQCDIR}. Here we give the relevant results in
dimensional regularization with $\overline{MS}$ renormalization
scheme, \bseq\label{NRQCDLDMEs}
\begin{align}
\label{1S0LDMEs}
\langle\cO(\cstate{1}{S}{0}{1})\rangle_{\state{1}{S}{0}}^{\mbox{{\footnotesize Born}}} \rightarrow
& \langle\cO(\cstate{1}{S}{0}{1})\rangle_{\state{1}{S}{0}}^{(\mu_{\Lambda})}\Big\{1-\frac{4}{3}\frac{\bq^{2}}{m_{Q}^{2}}
\frac{4\alpha_{s}}{3\pi}f_{\epsilon}(m_{Q})[\frac{1}{\epsilon}-\ln(\frac{\mu_{\Lambda}^{2}}{4m_Q^{2}})]\Big\},\\
\label{1P1LDMEs}
\begin{split}
\langle\cO(\cstate{1}{S}{0}{8})\rangle_{\state{1}{P}{1}}^{\mbox{{\footnotesize Born}}} \rightarrow
& \langle\cO(\cstate{1}{S}{0}{8})\rangle_{\state{1}{P}{1}}^{(\mu_{\Lambda})}\Big\{1-\frac{4}{3}\frac{\bq^{2}}{m_{Q}^{2}}
\frac{7\alpha_{s}}{12\pi}f_{\epsilon}(m_{Q})[\frac{1}{\epsilon}-\ln(\frac{\mu_{\Lambda}^{2}}{4m_Q^{2}})]\Big\}\\
&+\frac{16\alpha_{s}}{9\pi}f_{\epsilon}(m_{Q})\Big\{[\frac{1}{\epsilon}-\ln(\frac{\mu_{\Lambda}^{2}}{4m_Q^{2}})]\\
&+\frac{3\bq^{2}}{5m_{Q}^{2}}[-\frac{1}{\epsilon}+\ln(\frac{\mu_{\Lambda}^{2}}{4m_Q^{2}})]\Big\}
\frac{\langle\cO(\cstate{1}{P}{1}{1})\rangle_{\state{1}{P}{1}}^{\mbox{{\footnotesize Born}}}}{2N_{c}m_{Q}^{2}},
\end{split}\\
\label{P1P1LDMEs}
\langle\cP(\cstate{1}{S}{0}{8})\rangle_{\state{1}{P}{1}}^{\mbox{{\footnotesize Born}}}
\rightarrow& \langle\cP(\cstate{1}{S}{0}{8})\rangle_{\state{1}{P}{1}}^{(\mu_{\Lambda})}+
\frac{16\alpha_{s}}{9\pi}f_{\epsilon}(m_{Q})[\frac{1}{\epsilon}-\ln(\frac{\mu_{\Lambda}^{2}}{4m_Q^{2}})]
\frac{\langle\cP(\cstate{1}{P}{1}{1})\rangle_{\state{1}{P}{1}}^{\mbox{{\footnotesize Born}}}}{2N_{c}m_{Q}^{2}}
\end{align}
\eseq where $\mu_{\Lambda}$ is the factorization scale. Substituting
them into Eqs.~(\ref{Gamma1S0noNRQCD}) and ~(\ref{Gamma1P1noNRQCD}),
and considering the relation \bseq \bqa
\langle\cP(\cstate{1}{S}{0}{1})\rangle_{\state{1}{S}{0}}^{\mbox{{\footnotesize Born}}}
=& \bq^{2}\langle\cO(\cstate{1}{S}{0}{1})\rangle_{\state{1}{S}{0}}^{\mbox{{\footnotesize Born}}},\\
\langle\cP(\cstate{1}{S}{0}{8})\rangle_{\state{1}{P}{1}}^{\mbox{{\footnotesize Born}}}
=& \bq^{2}\langle\cO(\cstate{1}{S}{0}{8})\rangle_{\state{1}{P}{1}}^{\mbox{{\footnotesize Born}}},\\
\langle\cP(\cstate{1}{P}{1}{1})\rangle_{\state{1}{P}{1}}^{\mbox{{\footnotesize
Born}}} =&
\bq^{2}\langle\cO(\cstate{1}{P}{1}{1})\rangle_{\state{1}{P}{1}}^{\mbox{{\footnotesize
Born}}}, \eqa \eseq we get the SD coefficients for heavy quarkonium
hadronic decay of $S$-wave and $P$-wave states by matching full QCD and
NRQCD, \bseq \label{SDcoefficients}
\begin{align}
F(\cstate{1}{S}{0}{1}) =& \frac{4\pi\alpha_{s}^2}{9}\Big[1-\frac{\alpha_{s}}{\pi}
\frac{1}{72}(36\beta_{0}\ln(\frac{4m_{Q}^{2}}{\mu_{r}^{2}})+64n_{f}+93\pi^{2}-1908)\Big],\\
\begin{split}
G(\cstate{1}{S}{0}{1}) =& -\frac{4}{3}\frac{4\pi\alpha_{s}^2}{9}
\Big\{1-\frac{\alpha_{s}}{\pi}\frac{1}{144}[192\ln(\frac{\mu_{\Lambda}^{2}}{4m_Q^{2}})+72\beta_{0}\ln(\frac{4m_{Q}^{2}}{\mu_{r}^{2}})\label{G1S01}\\
&+164n_{f}+237\pi^{2}-4964]\Big\},
\end{split}\\
F(\cstate{1}{S}{0}{8}) =& \frac{5\pi\alpha_{s}^2}{6}\Big[1-\frac{\alpha_{s}}{\pi}
\frac{1}{72}(36\beta_{0}\ln(\frac{4m_{Q}^{2}}{\mu_{r}^{2}})+64n_{f}+129\pi^{2}-2448)\Big],\\
\begin{split}
G(\cstate{1}{S}{0}{8}) =& -\frac{4}{3}\frac{5\pi\alpha_{s}^2}{6}\Big\{1-\frac{\alpha_{s}}{\pi}
\frac{1}{288}[168\ln(\frac{\mu_{\Lambda}^{2}}{4m_Q^{2}})+144\beta_{0}\ln(\frac{4m_{Q}^{2}}{\mu_{r}^{2}})\\
&+328n_{f}+735\pi^{2}-12304]\Big\},
\end{split}\\
F(\cstate{1}{P}{1}{1}) =& \frac{5\alpha_{s}^{3}}{486}\Big[7(\pi^{2}-16)-24\ln(\frac{\mu_{\Lambda}^{2}}{4m_Q^{2}})\Big],\\
G(\cstate{1}{P}{1}{1}) =& \frac{\alpha_{s}^{3}}{3645}\Big[1740\ln(\frac{\mu_{\Lambda}^{2}}{4m_Q^{2}})-555\pi^{2}+9236\Big],
\end{align}
\eseq where $F$'s and $G$'s are defined in Eqs.~(\ref{1S0
factorization}) and (\ref{1P1 factorization}).

The SD coefficients of $\cstate{1}{S}{0}{1}$ agree with those in
Refs.~\cite{PhysRevD.51.1125,PhysRevD.66.094011,Petrelli:1997ge,Fan:2009cj,PhysRevD.83.114038},
that of $\cstate{1}{S}{0}{8}$ and $\cstate{1}{P}{1}{1}$ at leading
order in $v^2$ are also agree with previous results in
Ref.~\cite{Petrelli:1997ge}. The relativistic corrections
$G(\cstate{1}{S}{0}{8})$ and $G(\cstate{1}{P}{1}{1})$ are primarily
new results in this work. Based on these results, we will analyze
the decay of $\state{1}{S}{0}$ and $\state{1}{P}{1}$ heavy
quarkonium into light hadrons.

\section{PHENOMENOLOGICAL DISCUSSIONS}\label{sec:Phenomenology}
\subsection{Estimating NRQCD LDMEs}\label{sec:estimating}

To get the numerical result, we also need to know the value of
LDMEs. For $\state{1}{S}{0}$ quarkonium there are two LDMEs, and for
$\state{1}{P}{1}$ there are four. In Ref.~\cite{PhysRevD.83.114038}
the LDMEs of $\eta_{c}$ are determined by combining the Cornell
potential\cite{PhysRevD.17.3090} with one experimental measurement,
$\Gamma^{\mbox{{\footnotesize LH}}}(\eta_{c})$ or
$\Gamma^{\gamma\gamma}(\eta_{c})$\cite{Beringer:1900zz}, and
then one can predict other quantities. In the present work, since
there are not enough experimental inputs to determine all involved
LDMEs, we will estimate them by other methods.

For $\eta_{b}$, the situation is similar to
Ref.~\cite{PhysRevD.83.114038}, but lacking the experiment input of
the decay width to two photons $\Gamma^{\gamma\gamma}(\eta_{b})$. In
this case we will determine $\langle
\cO(\cstate{1}{S}{0}{1})\rangle_{\eta_{b}}$ from the potential model.
Here we use the Buchm\"uller-Tye(B-T) potential
model~\cite{PhysRevD.24.132} and Cornell(Corn) potential
model~\cite{PhysRevD.17.3090} results as input, which
give~\cite{PhysRevD.52.1726,Kang:2007uv} \bseq \bqa\label{Oetac}
\langle
\cO(\cstate{1}{S}{0}{1})\rangle_{\eta_{b}}^{\mbox{\footnotesize
B-T}}
&=& \frac{N_{c}}{2\pi}|R_{S}^{\mbox{\footnotesize B-T}}(0)|^{2}=3.093\,\mbox{GeV}^{3},\\
\langle
\cO(\cstate{1}{S}{0}{1})\rangle_{\eta_{b}}^{\mbox{\footnotesize
Corn}} &=& \langle
\cO(\cstate{3}{S}{1}{1})\rangle_{\Upsilon(1S)}^{\mbox{\footnotesize
Corn}}=3.07^{+0.21}_{-0.19}\,\mbox{GeV}^{3}. \label{Cornforetab}\eqa
\eseq In the Eq. (\ref{Cornforetab}) we use the heavy quark spin
symmetry(HQSS) to relate LDMEs of $\eta_b$ with that of
$\Upsilon(1S)$.
As the B-T model and Cornell model give
almost the same result, we will only use B-T model in the following.

In order to determine $\langle
\cP(\cstate{1}{S}{0}{1})\rangle_{\eta_{b}}$, we
define~\cite{PhysRevD.66.094011,PhysRevD.83.114038} \bqa \langle
\bm{v}^2 \rangle_{\eta_{b}} \equiv \frac{\langle
\cP(\cstate{1}{S}{0}{1})\rangle_{\eta_{b}}}{m_{b}^{2} \langle
\cO(\cstate{1}{S}{0}{1})\rangle_{\eta_{b}}}. \eqa
 Although $\langle
\bm{v}^2 \rangle_{\eta_{b}}$ can not be understood as the
expectation value of $\bm{v}^{2}$ in potential model, it can be
estimated from the Gremm-Kapustin relation \cite{Gremm:1997dq}
 \bqa \langle \bm{v}^2
\rangle_{\eta_{b}}^{\mbox{\footnotesize G-K}}=\frac{m_{\eta_{b}}-2
m_{pole}}{m_{pole}}. \eqa
Choosing $m_{pole}=4.6$ GeV for $b$
quark and $m_{\eta_{b}}=9.391$ GeV\cite{Beringer:1900zz}, we get $\langle \bm{v}^2 \rangle_{\eta_{b}}=0.042$,
which is close to the potential model estimated value $
\bm{v}^2\sim0.05-0.1$. Combining these results, we get the value of
redefined LDMEs in B-T model as \bqa \label{etabldmes}
\begin{split}
\langle\overline{\cO}(\cstate{1}{S}{0}{1})\rangle_{\eta_{b}} &\equiv
\frac{\langle\cO(\cstate{1}{S}{0}{1})\rangle_{\eta_{b}}}{2N_{c} m_{b}^{2}}=24.36^{+1.09}_{-1.03}\, {\text {MeV}},\\
\langle\overline{\cP}(\cstate{1}{S}{0}{1})\rangle_{\eta_{b}} &\equiv
\frac{\langle\cP(\cstate{1}{S}{0}{1})\rangle_{\eta_{b}}}{2N_{c}
m_{b}^{4}} =\langle\bm{v}^{2}\rangle_{\eta_{b}}
\langle\overline{\cO}(\cstate{1}{S}{0}{1})\rangle_{\eta_{b}}=1.01^{+0.05}_{-0.04}\,
\text{MeV}.
\end{split}
\eqa
 where the uncertainties are introduced by choosing $m_{b}=4.6\pm0.1$
GeV.

For $h_{c}$, we need to determine four LDMEs $\langle
\cO(\cstate{1}{P}{1}{1})\rangle_{h_{c}},\langle
\cO(\cstate{1}{S}{0}{8})\rangle_{h_{c}}, \langle
\cP(\cstate{1}{P}{1}{1})\rangle_{h_{c}}$ and $\langle
\cP(\cstate{1}{S}{0}{8})\rangle_{h_{c}}$. $\langle
\cO(\cstate{1}{P}{1}{1})\rangle_{h_{c}}$ is determined by the B-T
potential model~\cite{PhysRevD.52.1726} and \bqa\label{v2relations}
\langle \cP(\cstate{1}{P}{1}{1})\rangle_{h_{c}}\equiv \langle
\bm{v}^2 \rangle_{h_{c}} m_{c}^{2}\langle
\cO(\cstate{1}{P}{1}{1})\rangle_{h_{c}} \approx \langle \bm{v}^2
\rangle_{\eta_{c}} m_{c}^{2}\langle
\cO(\cstate{1}{P}{1}{1})\rangle_{h_{c}}, \eqa
 where $\langle \bm{v}^2
\rangle_{\eta_{c}}=0.228$ is taken from
Ref.~\cite{PhysRevD.83.114038}. Here we have tentatively assumed
$\langle \bm{v}^2 \rangle_{h_{c}}\approx\langle \bm{v}^2
\rangle_{\eta_{c}}$.
The remaining two color-octet LDMEs are determined by the operator
evolution method
(OEM)~\cite{PhysRevD.51.1125,Fan:2011aa,Gremm:1997dq}. From
Eq.~(\ref{eq:nrqcdNLO}) we get the evolution equations,
 \bqa \begin{split}\label{OERelations}
\mu_{\Lambda}^{2} \frac{d\langle\cO(\cstate{1}{S}{0}{8}) \rangle}{d \mu_{\Lambda}^{2}}
&= -\frac{7\alpha_{s}}{9\pi}\frac{\langle\cP(\cstate{1}{S}{0}{8}) \rangle}{m_{Q}^{2}}+
\frac{16\alpha_{s}}{9\pi}\frac{\langle\cO(\cstate{1}{P}{1}{1}) \rangle}{2 N_{c}m_{Q}^{2}}-
\frac{16\alpha_{s}}{15\pi}\frac{\langle\cP(\cstate{1}{P}{1}{1}) \rangle}{2 N_{c}m_{Q}^{4}},\\
\mu_{\Lambda}^{2} \frac{d\langle\cP(\cstate{1}{S}{0}{8}) \rangle}{d
\mu_{\Lambda}^{2}} &=
\frac{16\alpha_{s}}{9\pi}\frac{\langle\cP(\cstate{1}{P}{1}{1})
\rangle}{2 N_{c}m_{Q}^{2}}.\end{split} \eqa
 Knowing the values of $\langle\cO(\cstate{1}{P}{1}{1}) \rangle$ and
$\langle\cP(\cstate{1}{P}{1}{1}) \rangle$, the above differential
equations will determine the values of
$\langle\cO(\cstate{1}{S}{0}{8}) \rangle$ and
$\langle\cP(\cstate{1}{S}{0}{8}) \rangle$ by evolving from initial
values at $\mu_{\Lambda}=\mu_{\Lambda_0}$. Using two-loop running of
$\alpha_s$, we get
 \bqa \begin{split}
\langle\cO(\cstate{1}{S}{0}{8}) \rangle^{(\mu_\Lambda)}
&=\frac{64}{9 \beta_0}A\frac{\langle\cO(\cstate{1}{P}{1}{1})
\rangle}{2 N_{c}m_{Q}^{2}}-\frac{64}{3\beta_0}A\left(\frac{1}{5}+\frac{14}{27\beta_0}A\right)\frac{\langle\cP(\cstate{1}{P}{1}{1})
\rangle}{2 N_{c}m_{Q}^{4}}\\
&-\frac{28}{9\beta_0}A\frac{\langle\cP(\cstate{1}{S}{0}{8})\rangle^{(\mu_{\Lambda_0})}}{m_Q^2}+\langle\cO(\cstate{1}{S}{0}{8}) \rangle^{(\mu_{\Lambda_0})},\\
\langle\cP(\cstate{1}{S}{0}{8})\rangle^{(\mu_\Lambda)} &=
\frac{64}{9 \beta_0}A\frac{\langle\cP(\cstate{1}{P}{1}{1})
\rangle}{2
N_{c}m_{Q}^{2}}+\langle\cP(\cstate{1}{S}{0}{8})\rangle^{(\mu_{\Lambda_0})},\end{split}
\label{OEManalytic} \eqa where $\displaystyle
A\equiv\ln\frac{\alpha_s(\mu_{\Lambda0})}{\alpha_s(\mu_{\Lambda})}-\ln\frac{1+\alpha_s(\mu_{\Lambda0})\beta_1/\beta_0}{1+\alpha_s(\mu_{\Lambda})\beta_1/\beta_0}$
with $\beta_1=(17C_A^2-n_f T_R(10C_A+6C_F))/(6\pi)$. Choosing
$\mu_{\Lambda_0}= m_{c}v\sim 0.8\pm0.2$ GeV, the OEM assumes that
the the values of $\langle\cO(\cstate{1}{S}{0}{8}) \rangle$ and
$\langle\cP(\cstate{1}{S}{0}{8}) \rangle$ evaluated at
$\mu_{\Lambda} \approx2m_{c}$ can be estimated by the evolution term
only, i.e., neglecting initial values at $\mu_{\Lambda_0}$.
Set $m_c$ to be its pole mass, $1.5\pm0.1$
GeV, LDMEs at $\mu_{\Lambda}=2m_{c}$ are
 \bqa
\begin{split}\label{hcLDMEsnum1}
\langle \overline{\cO}(\cstate{1}{P}{1}{1})\rangle_{h_{c}} &\equiv
\frac{\langle \cO(\cstate{1}{P}{1}{1})\rangle_{h_{c}}}{2N_{c} m_{c}^4}=3.537_{-0.805}^{+1.124} \,\mbox{MeV},\\
\langle \overline{\cP}(\cstate{1}{P}{1}{1})\rangle_{h_{c}} &\equiv
\frac{\langle \cP(\cstate{1}{P}{1}{1})\rangle_{h_{c}}}{2N_{c} m_{c}^{6}}=0.806_{-0.183}^{+0.256} \,\mbox{MeV},\\
\langle \overline{\cO}(\cstate{1}{S}{0}{8})\rangle_{h_{c}} &\equiv
\frac{\langle \cO(\cstate{1}{S}{0}{8})\rangle_{h_{c}}}{m_{c}^{2}}=2.040_{-0.704}^{+1.208} \,\mbox{MeV},\\
\langle \overline{\cP}(\cstate{1}{S}{0}{8})\rangle_{h_{c}} &\equiv
\frac{\langle \cP(\cstate{1}{S}{0}{8})\rangle_{h_{c}}}{m_{c}^{4}}=0.561_{-0.197}^{+0.350} \,\mbox{MeV}.
\end{split}
\eqa The errors are estimated by varying $m_c$ and
$\mu_{\Lambda_0}$, among which, the uncertainty of $\mu_{\Lambda_0}$
dominates the errors for the two $S$-wave LDMEs.

Using the same method we can determine the LDMEs for $h_{b}$, \bqa
\begin{split}
&\langle
\overline{\cO}(\cstate{1}{P}{1}{1})\rangle_{h_{b}}=0.7555_{-0.0623}^{+0.0694}
\,
\mbox{MeV},\qquad \langle \overline{\cP}(\cstate{1}{P}{1}{1})\rangle_{h_{b}}=0.0314_{-0.0026}^{+0.0029} \,\mbox{MeV},\\
&\langle
\overline{\cO}(\cstate{1}{S}{0}{8})\rangle_{h_{b}}=0.3959_{-0.0503}^{+0.0611}
\,\mbox{MeV},\qquad \langle
\overline{\cP}(\cstate{1}{S}{0}{8})\rangle_{h_{b}}=0.0169_{-0.0022}^{+0.0026}
\,\mbox{MeV}.
\end{split}
\eqa
 Here we choose $m_{b}=4.6\pm0.1$ GeV, $\mu_{\Lambda_0}= m_{b}v\sim 1.5\pm0.2$ GeV and set $\langle
\bm{v}^2 \rangle_{h_{b}}\approx \langle \bm{v}^2
\rangle_{\eta_{b}}$, similar to the assumption for $h_{c}$.

Note that, another method to determine the value of the color-octet
LDME $\langle\cO(\cstate{1}{S}{0}{8})\rangle_{h_{c}}$ at leading
order in $v$ is provided in Ref.~\cite{Brambilla:2001xy}, where
LDMEs are further factorized by potential-NRQCD factorization, and
they are then expressed in terms of gluonic vacuum condensation
factor $\mathcal{E}(\mu)$. In Ref.~\cite{Brambilla:2001xy} they gave
both its evolution equation and the initial value at the scale
$\mu_0=1$GeV. We evolve this factor from the initial scale to
$2m_c$ and find that the value of
$\langle\overline{\cO}(\cstate{1}{S}{0}{8})\rangle_{h_{c}}$ through
this method is about 3.5 MeV, which is a little larger than our
result. However, the derivation is reasonable since we include the
relativistic corrections which essentially decrease the value at
leading order in $v$ [see the second term at the right-hand of the
first line in Eq.~(\ref{OEManalytic})].

\subsection{$\Gamma(\eta_{b}\rightarrow \mbox{LH})$}
We now discuss the hadronic decay width of $\eta_b$ based on the
values of LDMEs given above. Let's first fix both the
renormalization scale $\mu_{r}$ and factorization scale
$\mu_{\Lambda}$ to be $2 m_{b}$ and consider the uncertainty
introduced by LDMEs. For this choice of scales, the decay width can
be written as \bqa \Gamma(\eta_{b}\rightarrow
\mbox{LH})=427.4_{-5.7}^{+5.9}\times10^{-3}
\langle\overline{\cO}(\cstate{1}{S}{0}{1})\rangle_{\eta_{b}}-641.4_{-9.2}^{+8.8}\times10^{-3}
\langle\overline{\cP}(\cstate{1}{S}{0}{1})\rangle_{\eta_{b}} ,
\label{Gammaetabnum} \eqa
 where errors are estimated by varying
$m_b=4.6\pm0.1$\,GeV, and LDMEs
$\langle\overline{\cO}(\cstate{1}{S}{0}{1}\rangle_{\eta_{b}}$ and
$\langle\overline{\cP}(\cstate{1}{S}{0}{1}\rangle)_{\eta_{b}}$ are
given by Eq.~(\ref{etabldmes}). As the coefficients for
$\langle\overline{\cO}(\cstate{1}{S}{0}{1}\rangle_{\eta_{b}}$ and
$\langle\overline{\cP}(\cstate{1}{S}{0}{1}\rangle)_{\eta_{b}}$ are
at the same order, the smallness of
$\langle\overline{\cP}(\cstate{1}{S}{0}{1}\rangle)_{\eta_{b}}$ means
the relativistic correction can only change the total decay width by
about 5\%, which is not important as expected. Considering also the
correlation between errors, we get the hadronic decay width of
$\eta_{b}$ with the choice of $\mu_r=\mu_{\Lambda}=2m_b$, \bqa
\Gamma(\eta_{b}\rightarrow \mbox{LH})&= 9.76_{-0.54}^{+0.58}
\;\mbox{MeV}. \eqa
\begin{figure}
\includegraphics[scale=1]{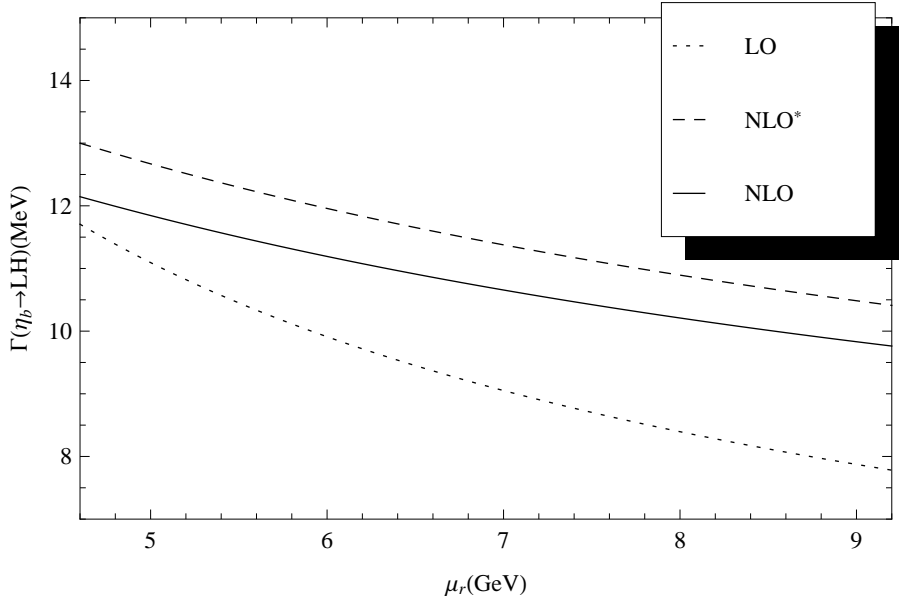}
\caption{\label{etabmurdep} $\mu_{r}$ dependence of
$\Gamma(\eta_{b}\rightarrow \mbox{LH})$. LO represents values
without QCD and relativistic corrections, NLO$^{*}$ includes QCD
corrections but only at leading order in $\bm{v}$, and NLO takes
into account all contributions up to $O(\alpha_{s}v^{2})$. The LDMEs
are taken from the B-T potential model and the Gremm-Kapustin
relation. Here we set $\mu_\Lambda=2 m_b$, and $m_b=4.6$\,GeV.}
\end{figure}
We find the $\mu_\Lambda$ dependence is much weaker than the $\mu_r$
dependence, thus we only discuss the $\mu_r$ dependence here. By
varying the $\mu_r$, we get the $\mu_r$ dependence of hadronic decay
width in FIG.~\ref{etabmurdep}. It is clear that the NLO calculation
significantly reduces the $\mu_r$ dependence.
Varying $\mu_r$ from $m_b$ to $2 m_b$, we get the decay width
$\Gamma(\eta_{b}\rightarrow \mbox{LH})\approx
\Gamma^{\mbox{\footnotesize total}}(\eta_{b})\sim 9.5-12$ MeV. This
value is consistent with the experimental data
$\Gamma^{\mbox{\footnotesize
exp}}(\eta_{b})=10.8^{+4.0}_{-3.7}{}^{+4.5}_{-2.0}$
MeV~\cite{Mizuk:2012pb}.

\subsection{$\Gamma(h_{c}\rightarrow \mbox{LH})$}
The numerical values of SD coefficients for hadronic decay width of
$h_{c}$ are \bqa
\begin{split}
\Gamma(h_{c}\rightarrow \mbox{LH})=& 328.7_{-21.8}^{+26.1}\times10^{-3} \langle
\overline{\cO}(\cstate{1}{S}{0}{8})\rangle_{h_{c}}-39.6_{-3.8}^{+3.1}\times10^{-3} \langle \overline{\cO}(\cstate{1}{P}{1}{1})\rangle_{h_{c}}\\
&-446.0_{-35.5}^{+29.7}\times10^{-3}\langle
\overline{\cP}(\cstate{1}{S}{0}{8})\rangle_{h_{c}}+92.4_{-7.3}^{+8.8}\times10^{-3} \langle
\overline{\cP}(\cstate{1}{P}{1}{1})\rangle_{h_{c}},
\end{split}
\eqa where both the renormalization scale $\mu_r$ and factorization
scale $\mu_\Lambda$ are set to be $2 m_c$. The redefined LDMEs and
their values are given in Eq.~(\ref{hcLDMEsnum1}). With these
results we then investigate the effects of the QCD corrections and
relativistic corrections.

\begin{figure}
\includegraphics[scale=0.65]{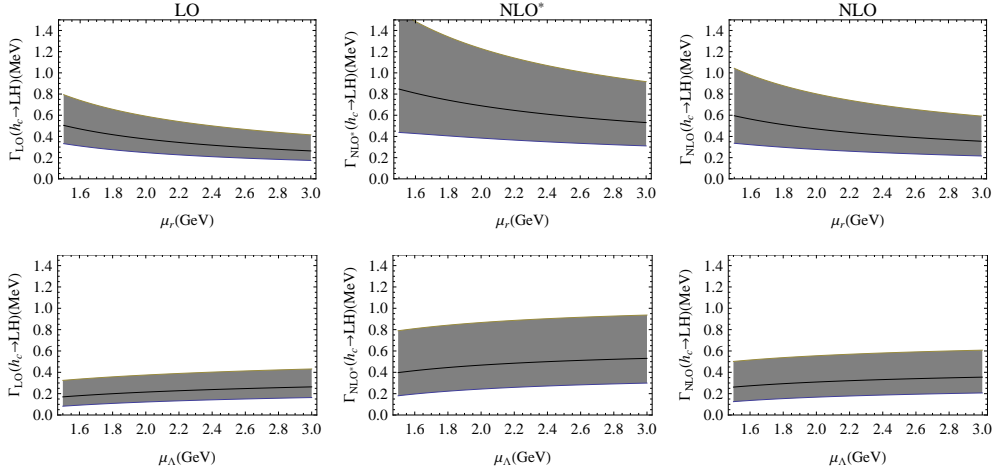}
\caption{\label{hcmudep} $\mu_{r}$ and $\mu_\Lambda$ dependence of
$\Gamma(h_{c}\rightarrow \mbox{LH})$. The upper plots are for
$\mu_{r}$ and lower ones for $\mu_\Lambda$. From left to right the
plots are shown for LO, NLO$^{*}$ and NLO  respectively, where
NLO$^{*}$ includes $O(\alpha_{s})$ but excludes $O(\alpha_{s}
v^{2})$ corrections.}
\end{figure}

Let us first analysis the partial widths of the four channels in
Table~\ref{table3}. Among the four, the
$\langle\cO(\cstate{1}{S}{0}{8})\rangle_{h_c}$ channel is positive
and it dominates the total width. Contributions of the
$\langle\cO(\cstate{1}{P}{1}{1})\rangle_{h_c}$ channel and
$\langle\cP(\cstate{1}{S}{0}{8})\rangle_{h_c}$  channel are negative
and compatible, although the latter one is suppressed by $v^2$. This
is because, as we mentioned before, the $\cstate{1}{P}{1}{1}$ Fock
state cannot couple with two gluons, and its SD coefficient is
suppressed by $\alpha_s$. It is the balance between $\alpha_s$ and $v^2$
that results in the two partial decay widths being compatible.
The last term,
$\langle\cP(\cstate{1}{P}{1}{1})\rangle_{h_c}$ channel, is
suppressed by both $\alpha_s$ and $v^2$, and it gives the smallest
contribution. Summing up the first two channels we get the decay
width at leading order in $v$,
$\Gamma^{(v^0)}=0.53^{+0.40}_{-0.23}$MeV, which is consistent with
the previous work~\cite{Maltoni:2000km}. However, we will show later
that the experimental data favor a smaller value. Including also the
relativistic corrections, the total decay width will decrease by
about $1/3$.
\begin{table}[ht]
\caption{ \label{table3} $\Gamma(h_{c}\rightarrow \mbox{LH})$
expressed with the contributions of each LDME.}
\begin{tabular}{|c|c|c|c|c|c|}
\hline
 & $\langle \cO(\cstate{1}{S}{0}{8})\rangle_{h_{c}}$ & $\langle \cO(\cstate{1}{P}{1}{1})
 \rangle_{h_{c}}$ & $\langle \cP(\cstate{1}{S}{0}{8})\rangle_{h_{c}}$ & $\langle \cP(\cstate{1}{P}{1}{1})\rangle_{h_{c}}$ & Total\\
\hline $\;\Gamma(\cstate{2S+1}{L}{J}{c}\rightarrow \mbox{LH})
\mbox{(MeV})\;$ & $0.67_{-0.25}^{+0.43}$ & $-0.14_{-0.06}^{+0.04}$ & $-0.25_{-0.17}^{+0.10}$ & $ 0.07_{-0.02}^{+0.03} $ & $\;0.35_{-0.15}^{+0.25}\;$\\
\hline
\end{tabular}
\end{table}
Next we list the partial widths order by order in $\alpha_s$ and $v$
in Table~\ref{table4}. We find the QCD correction, $\alpha_s^1 v^0$
contribution, is as large as the leading order contribution.
Detailed study reveals that the large correction mainly comes from
the $\cstate{1}{S}{0}{8}$ channel. In Ref.~\cite{Bodwin:2001pt}, the
authors pointed out that the large correction for
$\cstate{1}{S}{0}{1}$ channel, similar to the $\cstate{1}{S}{0}{8}$
channel, is due to the existence of renormalons, and they also
proposed a resummation method to deal with the renormalons. Nevertheless,
resummation of this kind for $\cstate{1}{S}{0}{8}$ channel is beyond
the scope of this work, and we will leave it as a future study. In our
work, as both of the $\alpha_s^0 v^2$ contribution and the
$\alpha_s^1 v^2$ contribution are negative, they can balance the
enhancement by QCD correction of $\cstate{1}{S}{0}{8}$ channel.
Moreover, we find our complete NLO correction improves the
normalization and factorization scale dependence compared with the
NLO* result, which are shown in FIG.~\ref{hcmudep}.

\begin{table}[ht]
\caption{ \label{table4} $\Gamma(h_{c}\rightarrow \mbox{LH})$
expressed with contributions at various orders of $\alpha_{s}$ and
$\bm{v}$.}.
\begin{tabular}{|c|c|c|c|c|c|}
\hline
 & $\alpha_{s}^{0}v^{0}$ & $\alpha_{s}^{1}v^{0}$ & $\alpha_{s}^{0}v^{2}$ & $\alpha_{s}^{1}v^{2}$ & Total\\
\hline $\;\Gamma(h_{c}\rightarrow \mbox{LH})
\mbox{(MeV})\;$& $0.32_{-0.12}^{+0.21}$ & $0.21_{-0.11}^{+0.20}$ & $-0.12_{-0.08}^{+0.04}$ & $ -0.06_{-0.08}^{+0.04} $ & $\;0.35_{-0.15}^{+0.25}\;$\\
\hline
\end{tabular}
\end{table}

In order to compare with the experiment data~\cite{Ablikim:2010rc},
we also need the E1 transition decay width $\Gamma(h_c\rightarrow
\eta_c +\gamma)$ up to the $v^2$ order, because this is another
important decay channel of $h_c$. Ref.~\cite{Maltoni:2000km}
estimated the transition decay widths but only at leading order in
$v$ by using HQSS between the spin-singlet and triplet P-wave
charmonia, \bqa \label{E1transition}
\Gamma(h_{c}\rightarrow\gamma\eta_c)=\frac{(E_{\gamma}^{h_c})^3}{9}\sum_{J=0}^{2}(2J+1)\frac{\Gamma(\chi_{cJ}\rightarrow\gamma
J/\psi)}{(E_\gamma^{\chi_{cJ}})^3}. \eqa And the obtained E1 width
is $615\pm29$ keV using the PDG Data~\cite{Beringer:1900zz}.
This result is consistent with the potential model calculations at
leading order in $v$~\cite{Chao:1992hd}. However, if the $v^2$
corrections are considered, HQSS will not hold any more.
Ref.~\cite{Chao:1992hd} showed that the width of
$h_{c}\rightarrow\gamma\eta_{c}$ can be reduced from 650~KeV to
385~KeV by relativistic effects. Subsequent studies using various
potential models~\cite{Gupta:1993pd,Godfrey:2002rp,Li:2009zu} also
observed similar relativistic effects, resulting in E1 transition
width at the range of 354-323~KeV. In this paper we choose the value
$\Gamma(h_c\rightarrow\gamma\eta_c)=385$ keV from
Ref.~\cite{Chao:1992hd}.

Combining the LH and $\gamma\eta_{c}$ decay channels of $h_c$, we
get the predictions for total decay width
$\Gamma^{\mbox{\footnotesize th}}(h_c)=0.74_{-0.15}^{+0.25}$ MeV and
the branching ratio $\mathcal{B}^{\mbox{\footnotesize
th}}(h_c\rightarrow \eta_{c}+\gamma)=52\pm13\%$. Our predictions are
consistent with the new experimental data $\Gamma^{\mbox{\footnotesize
exp}}(h_c)=0.73^{+0.45}_{-0.28}$ MeV and
$\mathcal{B}^{\mbox{\footnotesize
exp}}(h_{c}\rightarrow\eta_{c}+\gamma)=54.3\pm6.7\pm5.2\%$ measured by the BESIII Collaboration~\cite{Ablikim:2010rc}.
However, if we ignore the relativistic corrections to the hadronic
decay width, the total width will increase to 0.92 MeV and the E1
transition branching ratio will be decreased to $42\%$. Therefore,
it is evident that the relativistic corrections play an important
role in the $h_c$ decay and they can lead to a better agreement
between theoretical prediction and the experimental data.

\subsection{$\Gamma(h_{b}\rightarrow \mbox{LH})$}
\begin{figure}
    \includegraphics[scale=0.65]{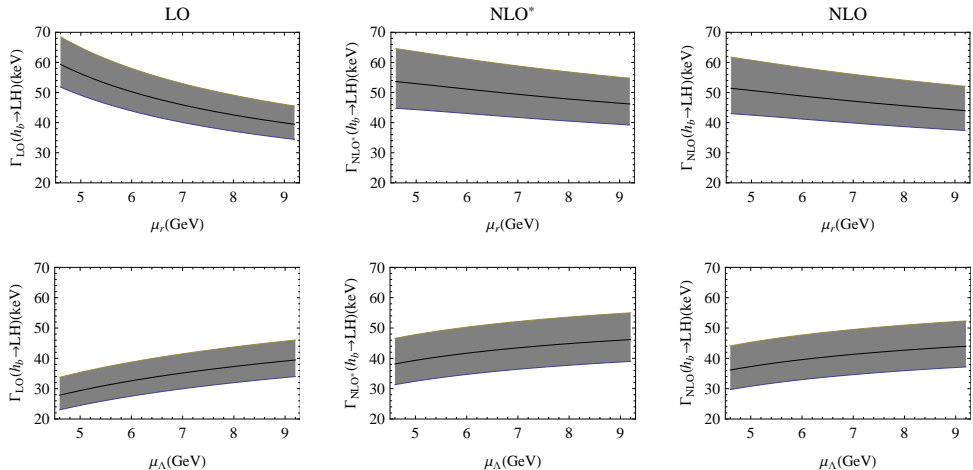}
\caption{\label{hbmudep} $\mu_{r}$ and $\mu_\Lambda$ dependence of
$\Gamma(h_{b}\rightarrow \mbox{LH})$. From left to right the three
plots represent LO, NLO$^{*}$ and NLO respectively, where NLO$^{*}$
includes $O(\alpha_{s})$ but excludes $O(\alpha_{s} v^{2})$
corrections. }
\end{figure}

Similar to $h_{c}$, we get the decay width for $h_{b}$, \bqa
\begin{split}
\Gamma(h_{b}\rightarrow \mbox{LH})=& 145.9_{-2.0}^{+2.1}\times10^{-3} \langle \overline{\cO}(
\cstate{1}{S}{0}{8})\rangle_{h_{b}}-(15.3\pm0.3)\times10^{-3} \langle \overline{\cO}(\cstate{1}{P}{1}{1})\rangle_{h_{b}}\\
&-(196.0\pm3.0)\times10^{-3}\langle
\overline{\cP}(\cstate{1}{S}{0}{8})\rangle_{h_{b}}+(35.8\pm0.6)\times10^{-3}
\langle \overline{\cP}(\cstate{1}{P}{1}{1})\rangle_{h_{b}}.
\end{split}
\eqa The $\mu_{r}$ and $\mu_\Lambda$ dependence are plotted in
Fig.~\ref{hbmudep}, where again we find the complete NLO correction
largely reduces the scale dependence. From partial decay width of
each contribution in Tables~\ref{table5} and~\ref{table6}, it is
clear that the $v^2$ correction effect is much smaller for $h_b$
than that for $h_c$, while QCD correction is still important. The E1
transition decay width for $h_b$ is evaluated in the
NR~\cite{Brambilla:2004wf}, GI~\cite{Godfrey:2002rp} and
Screened-potential models~\cite{Li:2009nr}, and the results are
listed in Table~\ref{table7}. Compared with the experiment data
$\mathcal{B}^{\mbox{\footnotesize
exp}}(h_b(1P)\rightarrow\eta_{b}(1S)\gamma)=49.2\pm5.7^{+5.6}_{-3.3}\%$~\cite{Mizuk:2012pb},
our prediction using NR model fits it very well, and predictions
using other three models are also within the error band.

\begin{table}[ht]
\caption{ \label{table5} $\Gamma(h_{b}\rightarrow \mbox{LH})$
expressed with contributions of each LDME.}
\begin{tabular}{|c|c|c|c|c|c|}
\hline
 & $\langle \cO(\cstate{1}{S}{0}{8})\rangle_{h_{b}}$ & $\langle \cO(\cstate{1}{P}{1}{1})
 \rangle_{h_{b}}$ & $\langle \cP(\cstate{1}{S}{0}{8})\rangle_{h_{b}}$ & $\langle \cP(\cstate{1}{P}{1}{1})\rangle_{h_{b}}$ & Total\\
\hline $\;\Gamma(\cstate{2S+1}{L}{J}{c}\rightarrow \mbox{LH})
\mbox{(keV})\;$ & $57.78_{-7.79}^{+9.42}$ & $-11.58_{-1.29}^{+1.13}$ & $-3.32_{-0.54}^{+0.45}$ & $ 1.12_{-0.11}^{+0.12} $ & $\;44.00_{-6.73}^{+8.23}\;$\\
\hline
\end{tabular}
\end{table}

\begin{table}[ht]
\caption{ \label{table6} $\Gamma(h_{b}\rightarrow \mbox{LH})$
expressed with various orders of $\alpha_{s}$ and $\bm{v}$.}.
\begin{tabular}{|c|c|c|c|c|c|}
\hline
 & $\alpha_{s}^{0}v^{0}$ & $\alpha_{s}^{1}v^{0}$ & $\alpha_{s}^{0}v^{2}$ & $\alpha_{s}^{1}v^{2}$ & Total\\
\hline $\;\Gamma(h_{b}\rightarrow \mbox{LH})
\mbox{(keV})\;$ & $33.41_{-4.46}^{+5.39}$ & $12.78_{-2.72}^{+3.39}$ & $-1.91_{-0.31}^{+0.26}$ & $ -0.29_{-0.19}^{+0.15} $ & $\;44.00_{-6.73}^{+8.23}\;$\\
\hline
\end{tabular}
\end{table}

\begin{table}[ht]
\caption{ \label{table7} $\Gamma(h_{b}\rightarrow \eta_b+\gamma)$
and $\mathcal{B}(h_{b}\rightarrow \eta_b+\gamma)$ in NR, GI and
Screened-potential models(SNR$_0$ is calculated using the
zeroth-order wave functions while SNR$_1$ using the first-order
relativistically corrected wave functions)}.
\begin{tabular}{|c|c|c|c|c|}
\hline
 & \quad NR \quad & \quad GI \quad & \quad SNR$_0$ \quad & \quad SNR$_1$ \quad\\
\hline
$\Gamma(h_{b}\rightarrow \eta_b+\gamma)$ (keV) & 41.8 & 37.0 & 55.8 & 36.3\\
\hline
$\Gamma_{\mbox{\footnotesize total}}(h_b)$ (keV) & 85.8 & 81.0 & 100.0 & 80.3\\
\hline
$\mathcal{B}(h_{b}\rightarrow \eta_b+\gamma)$  &48.7\% &  45.7\%& 55.9\% & 45.2\%\\
\hline
\end{tabular}
\end{table}


\section{SUMMARY}\label{sec:summary}

We have calculated order $\alpha_{s}v^{2}$ corrections for the
annihilation hadronic decay widths of spin-singlet heavy quarkonia
$\eta_{b}$, $h_{c}$ and $h_{b}$ within the framework of NRQCD. The
short-distance coefficients are calculated by covariant projection
method, and the LDMEs are estimated by using the potential model and
operator evolution methods. For the $h_{c}$ decay,
we find that $O(v^{2})$ and $O(\alpha_{s}v^{2})$ corrections
contribute  large and negative values to the decay width, which
substantially reduce the decay width calculated in the leading order
in $v^2$. It shows that relativistic corrections play an important
role in hadronic decays of $c\overline{c}$ system, and can improve
the theoretical results as compared with experimental data. Our
calculated total decay width $\Gamma^{\mbox{\footnotesize
th}}(h_c)=0.74_{-0.15}^{+0.25}$ MeV and branching ratio
$\mathcal{B}^{\mbox{\footnotesize th}}(h_c\rightarrow
\eta_{c}+\gamma)=52\pm13\%$ are consistent with the measurements by BESIII~\cite{Ablikim:2010rc}. For $\eta_{b}$ and $h_{b}$ decays, we
have calculated their hadronic decay widths and found that
$\Gamma(\eta_{b}\rightarrow \mbox{LH})=9.76_{-0.54}^{+0.58}$ MeV and
$\Gamma(h_{b}\rightarrow \mbox{LH})=44.00_{-6.73}^{+8.23}$ keV. We
conclude that for the $b\overline{b}$ system $O(\alpha_{s}v^{2})$
corrections are not as important as in the $c\overline{c}$ system.
We have also compared our theoretical results with experimental
data~\cite{Ablikim:2010rc,Mizuk:2012pb} and found that in general
our calculations are consistent with data within theoretical and
experimental uncertainties.

\begin{acknowledgments}
We are grateful to B.Q. Li, C. Meng, J.W. Qiu and M. Stratmann for
many helpful discussions. This work was supported in part by the
National Natural Science Foundation of China
(No.11021092 and No. 11075002), and the Ministry of Science and Technology
of China (No.2009CB825200). Y.Q.M is supported by the U.S.
Department of Energy, Contract No. DE-AC02- 98CH10886.
\end{acknowledgments}

\appendix
\section{EVOLUTION OF NRQCD MATRIX ELEMENTS
$\cO(\cstate{1}{S}{0}{8})$ and $\cP(\cstate{1}{S}{0}{8})$ AT
$O(\alpha_{s}v^{2})$}\label{NRQCDIR}

In order to cancel the infrared divergence in short-distance
coefficients of $\cstate{1}{P}{1}{1}$ Fock state, we need to
evaluate the NRQCD four-fermion operators $\cO(\cstate{1}{S}{0}{8})$
and $\cP(\cstate{1}{S}{0}{8})$ to sufficient order.

The $O(\alpha_{s})$ correction diagrams include three sets:
self-energy diagrams which are related to self-energy corrections of
external heavy (anti-)quarks; Coulomb diagrams where the gluon is
connected with both initial or final heavy quark and anti-quark; and
the intersecting diagrams where the gluon is related to an initial
heavy (anti-)quark and a final (anti-)quark. The results of the
first two sets have been given in
Refs.~\cite{Jia:2011ah,PhysRevD.83.114038}, and here we only
calculate the intersecting diagrams which relate to the transition
from $S$ wave to $P$ wave.

\begin{figure}
\includegraphics[scale=1.3]{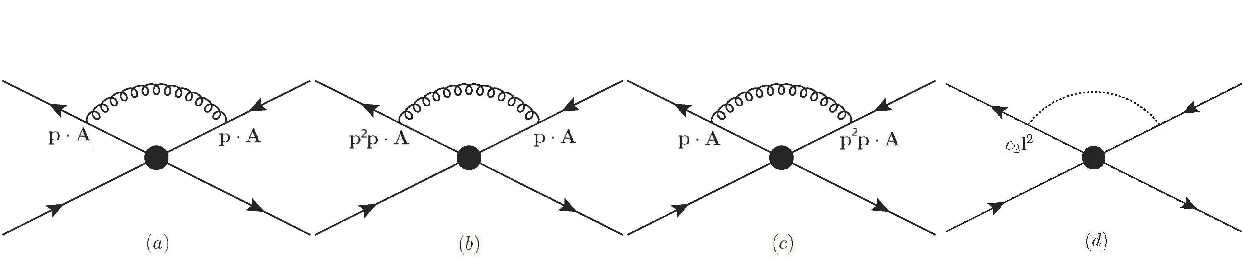}
\caption{\label{NRQCDv} The one-loop NRQCD diagrams which involve
the Feynman rules up to $O(v^2)$. The Coulomb interactions and the
cross diagrams have been suppressed.}
\end{figure}

Using the Lagrangian shown in Eqs.~(\ref{kineticLag}) and
(\ref{bilinearLag}), we can write the amplitudes of diagrams in
Fig.~\ref{NRQCDv} as (other crossed diagrams are not shown)

\bseq
\begin{align}
I_{a+b+c}&=i g_{s}^{2}\int\frac{d^{D}l}{(2\pi)^{D}}\frac{\bm{q}\mcdot\bm{q}^{\prime}-(\bm{q}\mcdot\bm{l})
(\bm{q}^{\prime}\mcdot\bm{l})/\bm{l}^{2}}{m_{Q}^{2}(l_{0}^{2}-\bm{l}^{2}+i\epsilon)}\frac{1-\bm{q}^{2}/2m_{Q}^{2}-\bm{q}^{\prime2}
/2m_{Q}^{2}}{[q_{0}-l_{0}-
\frac{(\bm{q}-\bm{l})^{2}}{2m_{Q}}+i\epsilon][q^{\prime}_{0}-l_{0}-
\frac{(\bm{q}^{\prime}-\bm{l})^{2}}{2m_{Q}}+i\epsilon]},\\
I_{d}&=i g_{s}^{2}\int\frac{d^{D}l}{(2\pi)^{D}}\frac{-1}{[q_{0}-l_{0}-
\frac{(\bm{q}-\bm{l})^{2}}{2m_{Q}}+i\epsilon][q^{\prime}_{0}-l_{0}-
\frac{(\bm{q}^{\prime}-\bm{l})^{2}}{2m_{Q}}+i\epsilon]},
\end{align}
\eseq where $q =(q_{0},\bm{q})$ is the heavy quark external momentum
and $l=(l_{0},\bm{l})$ is loop integral momentum. Since there is no
pole on the upper half of $l_{0}$'s complex plane, the second
integral $I_{d}$ yields zero. Contour integrating the first integral
over $l_{0}$ around the $l_{0}=|\bm{l}|-i\epsilon$ pole, we find
\begin{align}
I_{a+b+c}= g_{s}^{2}\int\frac{d^{D-1}l}{(2\pi)^{D-1}}\frac{\bm{q}\mcdot\bm{q}^{\prime}-
(\bm{q}\mcdot\bm{l})(\bm{q}^{\prime}\mcdot\bm{l})/\bm{l}^{2}}{2m_{Q}^{2}|\bm{l}|}\frac{1-\bm{q}^{2}/2m_{Q}^{2}-\bm{q}^{\prime2}/2m_{Q}^{2}}{[-|\bm{l}|
-\frac{\bm{l}^{2}}{2m_{Q}}+\frac{\bm{q}\cdot\bm{l}}{m_{Q}}+i\epsilon][-|\bm{l}|
-\frac{\bm{l}^{2}}{2m_{Q}}+\frac{\bm{q}^{\prime}\cdot\bm{l}}{m_{Q}}+i\epsilon]}.
\end{align}
Before further performing the integration, we will expand the
relative momentum in the denominator \cite{PhysRevD.55.2693}.
Assuming that $\bm{q}\mcdot\bm{l}/m_{Q}$,
$\bm{q}^{\prime}\mcdot\bm{l}/m_{Q}$ and $\bm{l}^2/m_{Q}$ are far
smaller than $|\bm{l}|$, we get the required expansion,
\begin{align}
I_{a+b+c}&= \frac{g_{s}^{2}}{2m_{Q}^{2}}\int\frac{d^{D-1}l}{(2\pi)^{D-1}}\frac{\bm{q}\mcdot\bm{q}^{\prime}-
(\bm{q}\mcdot\bm{l})(\bm{q}^{\prime}\mcdot\bm{l})/\bm{l}^{2}}{|\bm{l}|^{3}}(1-\bm{q}^{2}/2m_{Q}^{2}-\bm{q}^{\prime2}/2m_{Q}^{2})\nn\\
&\times\left(1+
(\frac{\bm{q}\mcdot\bm{l}}{|\bm{l}|m_{Q}})^{2}+(\frac{\bm{q}^{\prime}\mcdot\bm{l}}{|\bm{l}|m_{Q}})^{2}\right)+(\mbox{high order or irrelevant expansions}).
\end{align}
This integral can be reduced by taking the following substitution,
\bseq
\bqa
\bm{l}^{i}\bm{l}^{j} &\rightarrow& \frac{1}{D-1}\delta^{ij}\bm{l}^2,\\
\bm{l}^{i}\bm{l}^{j}\bm{l}^{k}\bm{l}^{r} &\rightarrow&
\frac{1}{(D-1)(D+1)}(\delta^{ij}\delta^{kr}+\delta^{ik}\delta^{jr}+\delta^{ir}\delta^{kj})\bm{l}^4,
\eqa \eseq where $\delta^{ij}$ is $D-1$ dimensional Euclidean delta
symbol. The integral yields
\bqa
I_{a+b+c}=\frac{\pi\alpha_{s}^{(b)}}{2
m_{Q}^2}\frac{\bm{q}\mcdot\bm{q}^{\prime}}{\pi^2}\frac{D-2}{D-1}(1-\frac{D-1}{D+1}
\frac{1}{2m_{Q}^2}(\bm{q}^{2}+\bm{q}^{\prime2}))\left(\frac{1}{\epsilon_{UV}}-\frac{1}{\epsilon_{IR}}\right).
\eqa
Summing up all the diagrams we get
\bqa
\begin{split}
I&=\frac{2 \alpha_{s}^{(b)}}{
\pi m_{Q}^2}\frac{D-2}{D-1} \bm{q}\mcdot\bm{q}^{\prime}\left(\frac{1}{\epsilon_{UV}}-
\frac{1}{\epsilon_{IR}}\right)(1-\frac{D-1}{D+1}\frac{1}{2m_{Q}^2}(\bm{q}^{2}+\bm{q}^{\prime2}))
\\ &\times\Big[C_{F}\frac{1\otimes 1}{2 N_{c}}+B_{F}T^{a}\otimes T^{a}\Big]\mathcal{O}(\cstate{1}{S}{0}{8}).
\end{split}
\eqa
Recalling the definitions of $\cO(\cstate{1}{P}{1}{1})$ and
$\cP(\cstate{1}{P}{1}{1})$, we can write \bseq\label{NRQCDloop}
\begin{align}
\begin{split}
\langle H|\cO(\cstate{1}{S}{0}{8})|H \rangle &= \langle H|\cO(\cstate{1}{S}{0}{8})|H
\rangle_{\mbox{\footnotesize Born}}+\frac{2(D-2)\alpha_{s}^{(b)}}{(D-1)\pi m_{Q}^2}\left(\frac{1}{\epsilon_{UV}}-\frac{1}{\epsilon_{IR}}\right)\\
&\times\left[C_{F}\frac{\langle H|\cO(\cstate{1}{P}{1}{1})|H \rangle}{2 N_{c}}-
\frac{D-1}{(D+1)m_{Q}^{2}}C_{F}\frac{\langle H|\cP(\cstate{1}{P}{1}{1})|H \rangle}{2 N_{c}}\right],
\end{split}\\
\langle H|\cP(\cstate{1}{S}{0}{8})|H \rangle &= \langle
H|\cP(\cstate{1}{S}{0}{8})|H \rangle_{\mbox{\footnotesize
Born}}+\frac{2(D-2)\alpha_{s}^{(b)}}{(D-1)\pi m_{Q}^2}
\left(\frac{1}{\epsilon_{UV}}-\frac{1}{\epsilon_{IR}}\right)C_{F}\frac{\langle
H|\cP(\cstate{1}{P}{1}{1}))|H \rangle}{2 N_{c}},
\end{align}
\eseq where we have omitted terms for $\cO(\cstate{1}{P}{1}{8})$ and
$\cP(\cstate{1}{P}{1}{8})$ since they are irrelevant in our work.
The presence of UV divergence indicates that the LDMEs need
renormalization. The relevant counter-term in the $\overline{MS}$
scheme can be chosen as \bseq\label{NRQCDcounterterm}
\begin{align}
\begin{split}
\langle H|\cO(\cstate{1}{S}{0}{8})|H \rangle &= \mu_{\Lambda}^{-2\epsilon}
\Bigg\{\langle H|\cO(\cstate{1}{S}{0}{8})|H \rangle^{(\mu_{\Lambda})}+
\frac{4\alpha_{s}}{3\pi m_{Q}^2}\left(\frac{1}{\epsilon_{UV}}+\ln4\pi-\gamma_{E}\right)\\
&\times\left[C_{F}\frac{\langle H|\cO(\cstate{1}{P}{1}{1})|H \rangle}{2 N_{c}}-
\frac{3}{5m_{Q}^{2}}C_{F}\frac{\langle H|\cP(\cstate{1}{P}{1}{1})|H \rangle}{2 N_{c}}\right]\Bigg\},
\end{split}\\
\begin{split}
\langle H|\cP(\cstate{1}{S}{0}{8})|H \rangle &=  \mu_{\Lambda}^{-2\epsilon}
\Bigg\{\langle H|\cP(\cstate{1}{S}{0}{8})|H \rangle^{(\mu_{\Lambda})}+
\frac{4\alpha_{s}}{3\pi m_{Q}^2}\left(\frac{1}{\epsilon_{UV}}+\ln4\pi-\gamma_{E}\right)\\
&\times C_{F}\frac{\langle H|\cP(\cstate{1}{P}{1}{1})|H \rangle}{2 N_{c}}\Bigg\},
\end{split}
\end{align}
\eseq where $\mu_{\Lambda}$ is the NRQCD renormalization scale.
Combining Eqs.~(\ref{NRQCDloop}) and~(\ref{NRQCDcounterterm}), we find
\bseq
\begin{align}
\begin{split}
\langle H|\cO(\cstate{1}{S}{0}{8})|H \rangle_{\mbox{\footnotesize Born}} &=
\mu_{\Lambda}^{-2\epsilon}\langle H|\cO(\cstate{1}{S}{0}{8})|H \rangle^{(\mu_{\Lambda})}+
\frac{4\alpha_{s}}{3\pi m_{Q}^2}\left(\frac{1}{\epsilon_{IR}}+\ln4\pi-\gamma_{E}\right)\\
&\times\left(\frac{\mu}{\mu_{\Lambda}}\right)^{2\epsilon}\left[C_{F}
\frac{\langle H|\cO(\cstate{1}{P}{1}{1})|H \rangle}{2 N_{c}}-\frac{3}{5m_{Q}^{2}}C_{F}\frac{\langle H|\cP(\cstate{1}{P}{1}{1})|H \rangle}{2 N_{c}}\right],
\end{split}\\
\begin{split}
\langle H|\cP(\cstate{1}{S}{0}{8})|H \rangle_{\mbox{\footnotesize Born}} &=
\mu_{\Lambda}^{-2\epsilon}\langle H|\cP(\cstate{1}{S}{0}{8})|H
\rangle^{(\mu_{\Lambda})}+\frac{4\alpha_{s}}{3\pi m_{Q}^2}\left(\frac{1}{\epsilon_{IR}}+\ln4\pi-\gamma_{E}\right)\\
&\times \left(\frac{\mu}{\mu_{\Lambda}}\right)^{2\epsilon}C_{F}\frac{\langle H|\cP(\cstate{1}{P}{1}{1})|H \rangle}{2 N_{c}}.
\end{split}
\end{align}
\eseq Considering also the self-energy contribution [see Eq. (B14)
in Ref.~\cite{PhysRevD.51.1125}], we get the total loop corrections
of NRQCD LDMEs, \bseq\label{eq:nrqcdNLO}
\begin{align}
\begin{split}
\langle H|\cO(\cstate{1}{S}{0}{8})|H \rangle_{\mbox{\footnotesize Born}} &=
\mu_{\Lambda}^{-2\epsilon}\langle H|\cO(\cstate{1}{S}{0}{8})|H \rangle^{(\mu_{\Lambda})}+
\frac{4\alpha_{s}}{3\pi m_{Q}^2}\left(\frac{1}{\epsilon_{IR}}+\ln4\pi-\gamma_{E}\right)\\
&\times\left(\frac{\mu}{\mu_{\Lambda}}\right)^{2\epsilon}\Bigg[C_{F}
\frac{\langle H|\cO(\cstate{1}{P}{1}{1})|H \rangle}{2 N_{c}}-\frac{3}{5m_{Q}^{2}}C_{F}\frac{\langle H|\cP(\cstate{1}{P}{1}{1})|H \rangle}{2 N_{c}}\\
&-\frac{N_{c}^{2}-2}{4N_{c}}\langle H|\cP(\cstate{1}{S}{0}{8}))|H \rangle\Bigg],
\end{split}\\
\begin{split}
\langle H|\cP(\cstate{1}{S}{0}{8})|H \rangle_{\mbox{\footnotesize Born}} &=
\mu_{\Lambda}^{-2\epsilon}\langle H|\cP(\cstate{1}{S}{0}{8})|H \rangle^{(\mu_{\Lambda})}+
\frac{4\alpha_{s}}{3\pi m_{Q}^2}\left(\frac{1}{\epsilon_{IR}}+\ln4\pi-\gamma_{E}\right)\\
&\times \left(\frac{\mu}{\mu_{\Lambda}}\right)^{2\epsilon}C_{F}\frac{\langle H|\cP(\cstate{1}{P}{1}{1})|H \rangle}{2 N_{c}}.
\end{split}
\end{align}
\eseq

\section{Scheme choice and absorption of $\langle
\mathcal{T}_{1-8}(\state{1}{S}{0},\state{1}{P}{1})\rangle_{\state{1}{P}{1}}$}\label{scheme}

In this appendix, we define the factorization scheme that we use in
this work, and we will show that there is no contribution from $\langle
\mathcal{T}_{1-8}(\state{1}{S}{0},\state{1}{P}{1})\rangle_{\state{1}{P}{1}}$
in our scheme. Let's begin with the factorization formula for
$\Gamma(H(\state{1}{P}{1})\rightarrow \textrm{LH})$ in $\msb$
scheme,
\begin{eqnarray}
\Gamma(H(\state{1}{P}{1})\rightarrow
\textrm{LH})&=&\frac{F(\cstate{1}{S}{0}{8})^\msb}{m_{Q}^{2}}\langle
\cO(\cstate{1}{S}{0}{8})\rangle_{\state{1}{P}{1}}^\msb+\frac{G(\cstate{1}{S}{0}{8})^\msb}{m_{Q}^{4}}\langle
\mathcal{P}(\cstate{1}{S}{0}{8})\rangle_{\state{1}{P}{1}}^\msb\nonumber\\
&+&\frac{F(\cstate{1}{P}{1}{1})^\msb}{m_{Q}^{4}}\langle
\cO(\cstate{1}{P}{1}{1})\rangle_{\state{1}{P}{1}}^\msb+
\frac{G(\cstate{1}{P}{1}{1})^\msb}{m_{Q}^{6}}\langle
\mathcal{P}(\cstate{1}{P}{1}{1})\rangle_{\state{1}{P}{1}}^\msb\nonumber\\
&+&\frac{T(\state{1}{S}{0},\state{1}{P}{1})^\msb}{m_{Q}^{5}}\langle
\mathcal{T}_{1-8}(\state{1}{S}{0},
\state{1}{P}{1})\rangle_{\state{1}{P}{1}}^\msb,\label{1P1msb}
\end{eqnarray}
where an explicit $\msb$ is marked for any LDME and SD coefficient.
There are many scheme choices to eliminate the last term in
Eq.~\eqref{1P1msb}. Our choice is to define the factorization scheme
of $\langle\cO(\cstate{1}{S}{0}{8})\rangle_{\state{1}{P}{1}}$ by the
following relation
\begin{eqnarray}
\Gamma(H(\state{1}{P}{1})\rightarrow
\textrm{LH})&=&\frac{F(\cstate{1}{S}{0}{8})^\msb}{m_{Q}^{2}}\langle
\cO(\cstate{1}{S}{0}{8})\rangle_{\state{1}{P}{1}}^{\text{LT}}+\frac{G(\cstate{1}{S}{0}{8})^\msb}{m_{Q}^{4}}\langle
\mathcal{P}(\cstate{1}{S}{0}{8})\rangle_{\state{1}{P}{1}}^\msb\nonumber\\
&+&\frac{F(\cstate{1}{P}{1}{1})^\msb}{m_{Q}^{4}}\langle
\cO(\cstate{1}{P}{1}{1})\rangle_{\state{1}{P}{1}}^\msb+
\frac{G(\cstate{1}{P}{1}{1})^\msb}{m_{Q}^{6}}\langle
\mathcal{P}(\cstate{1}{P}{1}{1})\rangle_{\state{1}{P}{1}}^\msb,\label{1P1clm}
\end{eqnarray}
where, to distinguish from $\msb$ scheme, we denote it as the leading
twist scheme (LT). Note that the relation in Eq.~\eqref{1P1clm}
should be understood to be valid only at $\alpha_s$ order, that is,
$T(\state{1}{S}{0},\state{1}{P}{1})^{\text{LT}}$ can be nonzero at
higher order in $\alpha_s$. From Eqs.~\eqref{1P1msb} and
\eqref{1P1clm}, we get the scheme transformation relation,
\begin{eqnarray}
\langle
\cO(\cstate{1}{S}{0}{8})\rangle_{\state{1}{P}{1}}^{\text{LT}}-\langle
\cO(\cstate{1}{S}{0}{8})\rangle_{\state{1}{P}{1}}^\msb=
\frac{T(\state{1}{S}{0},\state{1}{P}{1})^\msb}{m_{Q}^{3}
F(\cstate{1}{S}{0}{8})^\msb}\langle
\mathcal{T}_{1-8}(\state{1}{S}{0},
\state{1}{P}{1})\rangle_{\state{1}{P}{1}}^\msb.
\end{eqnarray}
According to the $\alpha_s$ expansion of SD coefficients,
\begin{subequations}
\begin{align}
F(\cstate{1}{S}{0}{8})^\msb=& F(\cstate{1}{S}{0}{8})^{(0)} + \alpha_s F(\cstate{1}{S}{0}{8})^{(1)\msb}+O(\alpha_s^2),\\
T(\state{1}{S}{0},\state{1}{P}{1})^\msb=& \alpha_s
T(\state{1}{S}{0},\state{1}{P}{1})^{(1)\msb}+O(\alpha_s^2),
\end{align}
\end{subequations}
we rewrite the difference as
\begin{eqnarray}\label{eq:clmdiff}
\langle
\cO(\cstate{1}{S}{0}{8})\rangle_{\state{1}{P}{1}}^{\text{LT}}-\langle
\cO(\cstate{1}{S}{0}{8})\rangle_{\state{1}{P}{1}}^\msb=\alpha_s
\frac{T(\state{1}{S}{0},\state{1}{P}{1})^{(1)\msb}}{m_{Q}^{3}
F(\cstate{1}{S}{0}{8})^{(0)}}\langle
\mathcal{T}_{1-8}(\state{1}{S}{0},
\state{1}{P}{1})\rangle_{\state{1}{P}{1}}^\msb+O(\alpha_s^2).
\end{eqnarray}
It is clear that the difference is suppressed by $\Oasv$,
Eq.~\eqref{1P1clm} does not determine the scheme choice of $\langle
\mathcal{T}_{1-8}(\state{1}{S}{0},
\state{1}{P}{1})\rangle_{\state{1}{P}{1}}$, and one can still choose
$\msb$ or other schemes. The reason is that the scheme dependence of
$\langle \mathcal{T}_{1-8}(\state{1}{S}{0},
\state{1}{P}{1})\rangle_{\state{1}{P}{1}}$ is at higher order in
$\alpha_s$, which is irrelevant to our calculation. Note that, the
relation between our scheme and $\msb$ scheme here is similar to the
relation between DIS scheme and $\msb$ scheme definition for the
$F_2$ structure function of virtual $\gamma$ deep inelastic
scattering (see Refs. \cite{Morfin:1990ck,Gluck:1994uf}, for
example).

An important consequence of Eq.~\eqref{eq:clmdiff} is that,
the evolution equations for $\langle
\cO(\cstate{1}{S}{0}{8})\rangle_{\state{1}{P}{1}}$ in both $\msb$
and LT scheme at $O(\alpha_s)$ are exactly the same, which follows
from the fact that the factorization scale dependence of both
$T(\state{1}{S}{0},\state{1}{P}{1})^{(1)\msb}$ and $\langle
\mathcal{T}_{1-8}(\state{1}{S}{0},
\state{1}{P}{1})\rangle_{\state{1}{P}{1}}^\msb$ are at
$O(\alpha_s)$. Therefore, although we calculate evolution equations
for LDMEs in $\msb$ scheme in Appendix \ref{NRQCDIR}, these results
are unchanged for the LT scheme.

Especially, the estimated $\langle
\cO(\cstate{1}{S}{0}{8})\rangle_{\state{1}{P}{1}}$  in
Sec.~\ref{sec:estimating} using OEM is the same for both LT scheme
and $\msb$ scheme. This seems to be questionable at first
glance, as Eq.~\eqref{eq:clmdiff} may imply its value is different
under the two different schemes. However, remember that the OEM
picks up only the evolution terms in the LDMEs and disregards all
other terms. Although Eq.~\eqref{eq:clmdiff} tells us that $\langle
\cO(\cstate{1}{S}{0}{8})\rangle_{\state{1}{P}{1}}$ is different
under the two schemes, the difference only changes the initial value,
which is ignored in the OEM. As a result, in the OEM this
difference is ignored.

\bibliographystyle{h-physrev}
\bibliography{NRQCDbib}


\end{document}